\documentclass[amsmath,twocolumn,aps]{revtex4-1} 

\usepackage{amsthm,amsfonts,graphicx,verbatim, xcolor}
\usepackage{hyperref}
\usepackage{bm}
\usepackage[utf8]{inputenc}
\usepackage{xfrac,dsfont}
\usepackage{kbordermatrix}

\let\oldmarginpar\marginpar
\renewcommand\marginpar[1]{\-\oldmarginpar[\raggedleft\footnotesize #1]%
{\raggedright\footnotesize #1}}
\newcommand{\bea}{\begin{eqnarray}}
\newcommand{\eea}{\end{eqnarray}}

\newcommand{\Tr}{{\rm Tr}\,}
\renewcommand{\epsilon}{\varepsilon}
\renewcommand{\vec}[1]{{\bf #1}}
\renewcommand{\Im}{{\rm Im}\,}
\renewcommand{\Re}{{\rm Re}\,}

\newcommand{\sign}{\mathrm{sign}}
\newcommand{\ket}[1]{|#1\rangle}
\newcommand{\bra}[1]{\langle #1|}
\newcommand{\braket}[2]{\langle #1|#2\rangle}
\newcommand{\braOket}[3]{\langle #1|#2|#3\rangle}

\def\matrix22#1#2#3#4{\left(\begin{array}{cc}#1&#2\\#3&#4\end{array}\right)}

\def\beq{\begin{equation}}
\def\eeq{\end{equation}}
\def\bea{\begin{eqnarray}}
\def\eea{\end{eqnarray}}

\begin{document}

\title{Matrix Product State Description and Gaplessness of the Haldane-Rezayi State}
\date{\today}
\author{Valentin Cr\'epel}
\affiliation{Laboratoire de Physique de l'\'Ecole normale sup\'erieure, ENS, Universit\'e PSL, CNRS, Sorbonne Universit\'e, Universit\'e Paris-Diderot, Sorbonne Paris Cit\'e, Paris, France.}
\author{Nicolas Regnault}
\affiliation{Laboratoire de Physique de l'\'Ecole normale sup\'erieure, ENS, Universit\'e PSL, CNRS, Sorbonne Universit\'e, Universit\'e Paris-Diderot, Sorbonne Paris Cit\'e, Paris, France.}
\author{Benoit Estienne}
\affiliation{Sorbonne Universit\'{e}, CNRS, Laboratoire de   Physique Th\'{e}orique et Hautes \'{E}nergies, LPTHE, F-75005 Paris, France.}

\begin{abstract}
	
	We derive an exact matrix product state representation of the Haldane-Rezayi state on both the cylinder and torus geometry. Our derivation is based on the description of the Haldane-Rezayi state as a correlator in a non-unitary logarithmic conformal field theory. This construction faithfully captures the ten degenerate ground states of this model state on the torus. Using the cylinder geometry, we probe the gapless nature of the phase by extracting the correlation length, which diverges in the thermodynamic limit. The numerically extracted topological entanglement entropies seem to only probe the Abelian part of the theory, which is reminiscent of the Gaffnian state, another model state deriving from a non-unitary conformal field theory.
	
\end{abstract}

\maketitle

\section{Introduction}

The success of the Laughlin ansatz~\cite{Laughlin_Ansatz} to describe a spinless Fractional Quantum Hall (FQH) system at filling $\nu = 1/3$ lies both in its predictive power through the plasma analogy~\cite{Laughlin_Plasma} and its microscopic relevance. Indeed, it is the densest zero energy state of a hollow-core Hamiltonian, the shortest repulsive interaction relevant for two spin-polarized fermions. In the Haldane's pseudopotential language~\cite{Haldane_HierarchiesPseudoPot}, the interaction correspond to the pure $V_1$ pseudo-potential penalizing any two fermions with angular momentum difference equal to one~\cite{Kivelson_PseudoPotentials}. Because the Laughlin wavefunction (WF) completely screens the largest pseudo-potential component of the Coulomb interaction projected in the lowest Landau Level (LL), it captures most of the features of the Ground-State (GS) of a system with repulsive Coulomb interactions. 

Applying the same reasoning to a spinful FQH system at filling $\nu=5/2$, Haldane and Rezayi proposed to approximate the Coulomb interaction with a pure $V_1$ pseudopotential, irrespective of the spin of the particles~\cite{HaldaneRezayi_Wavefunction}. Indeed, the contact interaction $V_0$, relevant for fermions with opposite spins, usually leads in magnitude in the lowest LL but is substantially reduced in the first LL. They argued that the plateau at $\nu=5/2$ can thus be described as a spinful system of $N_e$ electrons with $V_1$ SU(2)-symmetric interactions at filling $\nu=1/2$. They obtained the densest GS of this microscopic model, the so-called Haldane-Rezayi (HR) state. Despite these physical insights, the HR state shows some pathological behaviors. It exhibits a surprising ten-fold GS degeneracy on the torus~\cite{ReadRezayi_ZeroModes,Wen_TenFoldDegeneracy} and shows signs of criticality~\cite{Seidel_ThinTorusForHaldaneRezayi}. However, numerical studies in finite size with $N_e=8$ particles could not demonstrate the gapped or gapless nature of the phase~\cite{ReadRezayi_ZeroModes}.

The study of FQH model WFs greatly benefited from the insight of Moore and Read who realized that many of them could be written as Conformal Field Theory (CFT) correlators~\cite{MooreRead_CFTCorrelator}. This description relies on some assumptions like the gapped nature of the phase or the possibility to read off the universality class of the FQH state, the braiding and fusion properties of its low-energy excitations, from the bulk CFT. Testing these hypotheses and extracting physical observables such as the correlation length or the size of quasiparticles in the bulk cannot be done analytically from the conformal blocks and rely on numerical studies. A major progress to overcome the numerical bottleneck of these two-dimensional strongly interacting systems was made by Zaletel and Mong~\cite{ZaletelMong_MPS}. Going beyond the continuous MPS of Refs.~\cite{Dubail_EdgeStateInnerProduct} and~\cite{Sierra_iMPSCFT}, they realized that the CFT description of the states allows for an exact translation invariant and efficient Matrix Product State (MPS) description of these strongly correlated phases of matter. Combining the CFT construction with MPS algorithmic methods enables larger system sizes and predictions on physical observables previously out of reach~\cite{RegnaultEstienne_CorrelationLength,Regnault_PRLshort_MPSnonAbelianQH,Regnault_MPSnonAbelianQH,OurNatComms_H2L2,OurArxiv_MR2}.

The HR state can be expressed as a correlator within the $c=-2$ symplectic fermion CFT~\cite{MilovanovicRead_EdgePairedStates,Wen_TenFoldDegeneracy}. This non-unitary theory has negative scaling dimension operators, which are necessary to explain the ten-fold degeneracy of the GS manifold~\cite{GurarieFlohrNayak_HaldaneRezayiCFT}. Such a CFT cannot describe the edge physics of the system since the latter would then host unstable excitations with negative exponent correlations. Read provided strong arguments to show that non-unitarity generically implies bulk gaplessness~\cite{Read_NonUnitaryAndPpairing}. The lack of large-scale numerical evidence makes it hard to confirm or invalidate these theoretical predictions on the HR phase or to directly probe the physics of the hollow-core model.

In this article, we use an exact mapping of the symplectic fermion CFT to the $c=1$ Dirac CFT~\cite{Ludwig_RelatingNogCtoPosC,CapelliTodorov_PairedStates} to derive an easily implementable MPS describing the HR state and its zero energy quasihole excitations on the cylinder (Sec.~\ref{sec:HR_Construction}). We first use the transfer matrix formalism to show that the HR state has a diverging correlation length in the thermodynamic limit, convincingly proving the gaplessness of the hollow-core Hamiltonian (Sec.~\ref{sec:HR_Cylinder}). We adapt our MPS formulation to the torus geometry (Sec.~\ref{sec:MPS_Torus}), where a careful treatment of the zero modes allows us to recover the ten degenerate GS of the HR phase (Sec.~\ref{sec:HR_10fold}). They split into two groups. The first one is made of eight GS related by Abelian bulk excitations. Their topological entanglement entropy seems to only capture the Abelian part of the phase, tightly related to the Halperin 331 state. This feature is reminiscent of the Gaffnian state~\cite{Simon_Gaffnian}, also built on a non-unitary CFT, as shown in Ref.~\cite{RegnaultEstienne_CorrelationLength}. The second group consists of two states which appear as a Jordan block in the transfer matrix. They can be recovered with a twist operator located at the end of the system, which essentially plays the same role as the logarithmic operator in Ref.~\cite{GurarieFlohrNayak_HaldaneRezayiCFT}. Surprisingly, the topological entanglement entropy for the two states in the second group seems to be zero, similar to phases with trivial topological order such as the integer quantum Hall effect.

\section{The Haldane-Rezayi State and Conformal Field Theory} \label{sec:HR_Construction}

In this section, we give an overview of the HR state and summarize its properties. We then motivate our CFT description of the HR phase, and compare it to other works.

\subsection{Overview of the HR State} \label{ssec:Overview_HR}

The first quantized expression of the HR WF on genus zero surfaces such as the disk, the sphere or the cylinder is given by~\cite{HaldaneRezayi_Wavefunction}: \begin{equation} \label{eq:HaldaneRezayiFirstQuantized}
\begin{split}
\Psi^{\rm HR} (z_1, \cdots , z_N, & z_{[1]}, \cdots , z_{[N]}) =  \\ 
& \det\left[ \frac{1}{(z_i-z_{[j]})^2} \right] \prod_{i<j}^{2N} (z_i-z_j)^2 \, .
\end{split}
\end{equation} Here, $z_i$ (resp. $z_{[i]}$) denotes the position of the $i$-th spin up (resp. down) particle, bracketed indices are identified as $[i]=i+N$ in the last product. We omit the LL Gaussian measure. As stated in the introduction, the HR state of Eq.~\ref{eq:HaldaneRezayiFirstQuantized} is the densest zero energy state of a system of $N_e=2N$ spin-\sfrac{1}{2} electrons in the Lowest LL (LLL) which, irrespective of the spin, interact through a $V_1$ two-body pseudopotential
\begin{equation} \label{eq:V1_PseudoPot}
V_1 (\vec{r}= \vec{r}_1 - \vec{r}_2) = - \nabla_{\vec{r}}^2 \, \delta^{(2)}(\vec{r}) \, ,
\end{equation} where $\vec{r}$ is the relative position of the two particles. On the sphere, this unique densest zero energy state occurs when the number of flux quanta $N_\phi$ satisfies $N_\phi = 2N_e-4$. This hollow-core interacting Hamiltonian hosts many more, albeit less dense, zero energy states corresponding to edge and bulk quasihole excitations of Eq.~\ref{eq:HaldaneRezayiFirstQuantized}. The different types of bulk quasiparticles, their charge and braiding properties encode the topological content of the phase.

Decoupling of the spin and charge degrees of freedom, expected for the system one-dimensional edge effective description~\cite{Voit_SpinCharge}, also occurs in the bulk as can be seen from the factorization of the HR state into a determinant, encoding the $d$-wave pairing of the electrons into a spin-singlet~\cite{ReadGreen_PariedFermions}, and a Jastrow factor which is associated with the electric charge~\cite{MooreRead_CFTCorrelator}. The latter sets the filling fraction to $\nu=1/2$ and describes the Laughlin-like $e/2$ Abelian quasiholes of the system~\cite{Fubini_VertexOperatorsFQHE}. These excitations do not affect the pairing part of the WF. The treatment of the spin degrees of freedom requires a more careful treatment. It was shown in Ref.~\cite{ReadRezayi_ZeroModes} that there are $2^{2n-3}$ linearly independent zero-energy states of the hollow-core model Hamiltonian with $2n$ neutral quasiholes at given positions. This exponential degeneracy often evidences non-Abelian statistics~\cite{Wilczek_FractionalSpinAnyon,NayakWilczek_SpinorBraiding}. However, the statistics can only be infered after identification of the distinct neutral quasihole excitations.

For that purpose, it is useful to consider the system on a torus. Indeed, on this geometry the GS degeneracy equals the number of distinct bulk excitations that the Hall state admits. The hollow-core Hamiltonian of Eq.~\ref{eq:V1_PseudoPot} has ten degenerate GS on the torus~\cite{Wen_TenFoldDegeneracy}, which split into two halves depending on the $e/2$ quasihole parity~\cite{HermannsRegnaultArdonne_Nonunitaries}. Neutral excitations are thus responsible for a five-fold degeneracy. The first quantized expressions of the corresponding WFs on the torus were obtained in Ref.~\cite{ReadRezayi_ZeroModes}. Four of the GS are generalizations of Eq.~\ref{eq:HaldaneRezayiFirstQuantized} on the torus. The fifth state has two unpaired electrons, and is totally antisymmetric over all possible way to choose these two electrons. This introduces some long-range behaviors in the WFs, as observed in Ref.~\cite{Seidel_ThinTorusForHaldaneRezayi}. The quasiparticle relating any of the four first GS to the fifth one creates these long-range correlations. Such a feature can be captured using the non-unitary symplectic fermion CFT to describe the neutral degrees of freedom~\cite{MilovanovicRead_EdgePairedStates}, in which the negative conformal dimension operators induces these non-local correlations. This theory furthermore support the non-Abelian nature of the phase~\cite{GurarieFlohrNayak_HaldaneRezayiCFT}.

What are the physical consequence of this non-unitary neutral CFT? Read provided compelling arguments in Ref.~\cite{Read_NonUnitaryAndPpairing} that FQH model WFs built on a non-unitary CFT generically describe compressible states, and thus could not capture the physics of a Hall quantized conductance plateau. For instance the Gaffnian~\cite{Simon_Gaffnian}, which relies on a non-unitary CFT, was shown to be critical~\cite{RegnaultEstienne_CorrelationLength,JolicoeurLechemiant_GaplessHR}. The HR state is suspected to follow the same behavior~\cite{ReadGreen_PariedFermions,Seidel_ThinTorusForHaldaneRezayi}. At the edge of the system, a negative scaling dimension leads to an unstable theory. It is however possible that at the edge, some correction to the symplectic fermions stress energy tensor stabilizes another unitary theory~\cite{GurarieFlohrNayak_HaldaneRezayiCFT}. The latter should have the same characters as the non-unitary CFT, which are completely determined by the construction of all zero energy states of the hollow-core model~\cite{MilovanovicRead_EdgePairedStates}.

\subsection{CFT Description} \label{ssec:OurCFTDescription}

The Jastrow factor of Eq.~\ref{eq:HaldaneRezayiFirstQuantized} is associated with the electric charge whose degrees of freedom are described by a free massless chiral boson $\varphi(z)$ compactified on a circle of radius $R=\sqrt{2}$ (see Ref.~\cite{YellowBook} for a review). An important primary field of the theory is the vertex operator 
\begin{equation}
\mathbf{V}^c (z) = : e^{i \sqrt{2} \varphi (z)} : \, 
\end{equation} whose correlator reproduces the Jastrow factor:
\begin{equation}
\langle \mathcal{O}_{\rm Bkg} \mathbf{V}^c (z_1) \cdots \mathbf{V}^c (z_{[N]}) \rangle = \prod_{i=1}^{2N} (z_i-z_j)^2 \, .
\end{equation} The neutralizing background charge $\mathcal{O}_{\rm Bkg} = \exp \left(-i2\sqrt{2}N \varphi_0 \right)$, with $\varphi_0$ the bosonic zero-point momentum, is inserted to make the correlator non-vanishing~\cite{YellowBook,MooreRead_CFTCorrelator}. This is the usual treatment of Jastrow-like FQH model states~\cite{Fubini_VertexOperatorsFQHE}. It sets the filling fraction to $\nu=1/2$ and describes the bulk $e/2$ Abelian quasiholes of the system.

The neutral part of Eq.~\ref{eq:HaldaneRezayiFirstQuantized} can be reproduced as the $2N$-point correlator of a pair of free fields $\Psi^\uparrow$ and $\Psi^\downarrow$ \begin{equation} \label{eq:NeutralPartAsCorrelator_General}
\langle \Psi^\uparrow (z_1) \Psi^\downarrow (z_{[1]}) \cdots \Psi^\uparrow (z_N) \Psi^\downarrow (z_{[N]}) \rangle = \det\left[ \frac{1}{(z_i-z_{[j]})^2} \right]  \, .
\end{equation} For Eq.~\ref{eq:NeutralPartAsCorrelator_General} to hold, it is sufficient that the fields obey the fermionic Wick theorem, and that their two-point correlation function reads $\langle \Psi^\sigma (z) \Psi^\rho (0)\rangle = \varepsilon^{\rho \sigma}/z^2$. These conditions are for instance realized in the free fermionic models of Refs.~\cite{LeeWen_HaldaneRezayiWithoutCFT} and~\cite{MilovanovicRead_EdgePairedStates} or in the non-unitary $c=-2$ symplectic fermions CFT~\cite{GurarieFlohrNayak_HaldaneRezayiCFT}. The latter uses logarithmic operators to predict a ten-fold GS degeneracy on the torus, matching the numerically observed number of GS using ED~\cite{HermannsRegnaultArdonne_Nonunitaries}. The braiding statistics of quasi-holes inferred from this non-unitary theory substantiate the possible non-Abelian nature of the HR phase. The construction of all zero energy states of the hollow core Hamiltonian is enough to specify the characters of the edge theory but it does not fix the stress-energy tensor of the latter. Hence, we may find other ways to describe the neutral degrees of freedom of the HR phase. It is known~\cite{CapelliTodorov_PairedStates} how to change the action of the $c=-2$ symplectic fermions (equivalent to a $(\xi, \eta)$ conformal weight (0,1) fermionic ghost sysem~\cite{YellowBook}) to reach the unitary Dirac fermion $(\psi,\psi^\dagger)$ CFT (realized as conformal weight (\sfrac{1}{2},\sfrac{1}{2}) fermionic ghosts~\cite{FlohrLectureLogarithmicCFT}), without changing the characters of the edge theory~\cite{Ludwig_RelatingNogCtoPosC}. 

In this article, we choose to represent the neutral degrees of freedom of the HR state with a $c=1$ Dirac CFT. Differentiation of the $2N$-point correlator $\langle \psi(z_1) \psi^\dagger(z_{[1]}) \cdots \psi(z_N) \psi^\dagger(z_{[N]}) \rangle = \det \left[(z_i - z_{[j]})^{-1}\right]$~\cite{Ginspard_CFTlectures} shows that the fields
\begin{equation} \label{eq:Definition_SpinFields}
\Psi^\uparrow = \psi^\dagger \quad \text{and} \quad \Psi^\downarrow = \partial \psi \, ,
\end{equation} satisfy Eq.~\ref{eq:NeutralPartAsCorrelator_General}.

Combining the bosonic and fermionic parts, the full HR bulk WF is obtained as the correlator of the following electronic operators 
\begin{equation} \label{eq:Full_ElecOperators}
\begin{split}
\mathcal{V}^\uparrow (z) = \Psi^\uparrow (z) \cdot \mathbf{V}^c(z) \, , \\  \mathcal{V}^\downarrow (z) = \Psi^\downarrow (z) \cdot \mathbf{V}^c(z) \, . 
\end{split}
\end{equation} This construction is closely related to the CFT description of the Halperin 331 state~\cite{Halperin_Ansatz} via the bosonization identities $\psi^\dagger = :e^{i\varphi_s}:$ and $\psi =:e^{-i\varphi_s}:$, where $\varphi_s$ is a chiral massless boson with unit compactification radius, encoding the spin degrees of freedom. Here, only the derivative in Eq.~\ref{eq:Definition_SpinFields} differs from the Halperin construction~\cite{OurPaper_HalperinMPS}. The writing of Eq.~\ref{eq:Full_ElecOperators} based on $(\varphi, \varphi_s)$ exactly matches the CFT description of a level-2 hierarchy state as given in Ref.~\cite{HierarchiesCFT_Review}. The hierarchy WF describes a spinful state at filling $\nu=1/2$, which results from quasielectron condensation~\cite{HanssonHermanns_QuasielectronCFT}. In this context, the derivatives emerge when regularizing the OPE between electrons and quasi-particle operators~\cite{SuorsaViefersHansson_GeneralApproachHallHierarchies}. We thus expect, as shown in Refs.~\cite{SuorsaViefersHansson_GeneralApproachHallHierarchies} and~\cite{ViefersHansson_TaoThoulessLimitCFT}, that the hierarchy state is consistent with the \textbf{K}-matrix classification~\cite{WenZee_Kmatrix}, with
\begin{equation} \label{eq:Kmatrix_331}
\textbf{K} = \begin{pmatrix} 3 & 1 \\ 1 & 3 \end{pmatrix} \, ,
\end{equation} 
and where the derivative adds one unit to the conformal spin of $\mathcal{V}^\downarrow$. This connection might seem strange at first sight because the hierarchy construction produces an Abelian phase with $|\det \textbf{K}| = 8$ degenerate GS on the torus~\cite{HierarchiesCFT_Review}. We will elucidate the difference between the two theories by a careful treatment of the zero modes in Sec.~\ref{sec:HR_10fold}.

\subsection{CFT Hilbert Space} \label{ssec:CFT_HilbertSpace}

In the following, we will often rely on the cylinder geometry with coordinate $w = x + iy$ obtained from the plane through the conformal transformation $z = \exp \left( \gamma (x+iy) \right)$ with $\gamma = \frac{2\pi}{L}$. $x$ denotes the coordinate along the cylinder axis, $y$ being that along the compact dimension. We assume periodic boundary conditions for the electronic operators Eq.~\ref{eq:Full_ElecOperators} when they wind around the cylinder. The CFT thus splits into two parts \textbf{P} and \textbf{AP} in which $\Psi^\uparrow$, $\Psi^\downarrow$ and $\mathbf{V}^c$ are respectively periodic with integer modes and anti-periodic with half-integer modes. As a consequence, the electronic modes $\mathcal{V}^\uparrow$ and $\mathcal{V}^\downarrow$ can be computed from Eq.~\ref{eq:Definition_SpinFields} and Eq.~\ref{eq:Full_ElecOperators}:
\begin{equation} \label{eq:ModeExpansion_ElecOperators}
\begin{split}
\mathcal{V}_{-\lambda}^\uparrow & = \sum_{n} \psi_{n}^\dagger \cdot \mathbf{V}_{-n-\lambda}^c \, , \\  \mathcal{V}_{-\lambda}^\downarrow & = \sum_{n} (-n) \, \psi_n \cdot  \mathbf{V}_{-n-\lambda}^c \, ,
\end{split}
\end{equation} where the boundary conditions require $\lambda \in \mathbb{Z}$ in both sectors while $n \in \mathbb{Z}$ in \textbf{P} and $n \in \mathbb{Z}+1/2$ in \textbf{AP}. Here and thereafter, we denote as $\phi_{n}$ the $n$-th mode of a primary field $\phi = \sum_n e^{- \gamma n w} \, \phi_{n} $ on the cylinder.

On this geometry, the free boson has the following mode expansion:
\begin{equation} \label{eq:FreeBoson}
\varphi (w) = \varphi_0 - i \gamma w a_0  + i \sum_{n\in \mathbb{Z}^*} \dfrac{1}{n} \, e^{ - \gamma n w } \, a_{n} . 
\end{equation} The U(1) Kac-Moody algebra satisfied by the bosonic mode, $[a_n ; a_m]=n \delta_{m+n,0}$, implies the electric charge conservation through the conserved current $J(z)=i\partial \varphi (z)$. The U(1)-charge, measured in units of half the electron charge $e/2$ by $R a_0 = \sqrt{2} a_0$, must be either integer in \textbf{P} or half-integer in \textbf{AP}. The zero point momentum $\varphi_0$ is the canonical conjugate of $a_0$, \textit{i.e.} $[\varphi_0,a_0]=i$. As such, the operator 
\begin{equation} \label{eq:BackgroundCharge_Charge}
U_c = e^{-(i/R) \varphi_0}
\end{equation}
removes one unit of charge. Primary states of the bosonic CFT Hilbert space, labeled by their U(1)-charge, are obtained as $\ket{q} = \lim\limits_{y \to -\infty} e^{-(q/R) \varphi (w)} \ket{0}$ with $\ket{0}$ being the bosonic CFT vacuum. The Operator Product Expansion (OPE) between vertex operators $\mathbf{V}^c (z) \ket{q} \sim z^q \ket{q+2} + \cdots$~\cite{YellowBook} ensures the correct boundary conditions for $\mathbf{V}^c$ thanks to the charge selection rules in \textbf{P} and \textbf{AP}. 

The fermionic modes (integer in \textbf{P} and half integer in \textbf{AP}) satisfy the anticommutation relations $\{\psi_{n} , \psi_{m}^\dagger \} = \delta_{m+n,0}$. In the periodic sector, the zero modes anticommutation relations lead to a set of degenerate highest-weight states $\{\ket{\sigma_i}\}_i$~\cite{Ginspard_CFTlectures}. They physically correspond to modes precisely at the Fermi energy which can either be occupied or non-occupied. They are obtained by acting with twist operators of dimension $1/8$ on the fermionic vacuum $\ket{\mathbb{I}}$. 

We can build the full CFT Hilbert space, which is also the virtual (or auxiliary) space of our MPS description, from the bosonic and fermionic ones. It is obtained by repeated action of the creation operators $a_{-n}$, $\psi_{-n}$ and $\psi_{-n}^\dagger$ with $n\in \mathbb{N}^*$ on the highest weight states compatible with the boundary condition. These actions are encoded in one bosonic partition $\mu$ and two fermionic ones $(\eta, \nu)$:
\begin{equation} \label{eq:CFT_Hilbert}
\begin{split}
& \textbf{AP}:  \, \ket{q, \mathbb{I}, \mu, \eta, \nu} = \prod_{i\in \mathbb{N}} a_{-\mu_i} \psi_{-\eta_i} \psi_{-\nu_i}^\dagger (\ket{q} \otimes \ket{\mathbb{I}}) \, , \\ & \textbf{P}:  \,   \ket{q, \sigma_i, \mu, \eta, \nu} = \prod_{i\in \mathbb{N}} a_{-\mu_i} \psi_{-\eta_i} \psi_{-\nu_i}^\dagger (\ket{q} \otimes \ket{\sigma_i}) \, .
\end{split}
\end{equation} Here, $(\eta_i, \nu_i) \in \mathbb{N}^*$ in \textbf{P} and $(\eta_i, \nu_i) \in \mathbb{N}+1/2$ in \textbf{AP} are the non-repeated elements of the fermionic  partitions $\eta$ and $\nu$. The bosonic degrees of freedom are described with the bosonic partition $\mu$ whose possibly repeated elements are $\mu_i \in \mathbb{N}^*$, and a U(1)-charge constrained to be $q \in \mathbb{Z}$ in \textbf{P} and half-integer $q \in \mathbb{Z}+1/2$ in \textbf{AP}. The CFT space divides into four charge sectors which are stable under the action of the electronic operators~\cite{ReadRezayi_ZeroModes,MilovanovicRead_EdgePairedStates}. We label them with $a \in \{ 0, 1/2, 1, 3/2 \}$, each of these four sectors gathers all the states with U(1)-charge equal to $a$ mod 2 (physically, modulo the elementary charge $e$). The two sectors in \textbf{P} ($a=0$ and $a=1$) or \textbf{AP} ($a=1/2$ and $a=3/2$) are related by a unit shift of the bosonic charge, which corresponds to a center of mass translation on the cylinder, and they thus share the same physical properties. Each of the $a$ sectors splits into two depending on the number of fermions in Eq.~\ref{eq:CFT_Hilbert} leading to the total eight hierarchy-like topological sectors discussed in Sec.~\ref{ssec:Overview_HR}~\cite{OurPaper_HalperinMPS}.

\subsection{Relation to Other Approaches}

Our CFT description of the HR electronic operators Eq.~\ref{eq:Definition_SpinFields} agrees with that of Ref.~\cite{CapelliTodorov_PairedStates}. As previously mentioned, a more common approach~\cite{Jolicoeur_ModifiedCoulombGas,Yoshioka_PairingSymEvenDenom,Flohr_BCforJainStates} relies on the non-unitary $c=-2$ CFT as first described in Ref.~\cite{GurarieFlohrNayak_HaldaneRezayiCFT}. Using an exact mapping between the $c=-2$ symplectic fermion theory and the the Dirac CFT proposed by Guruswamy and Ludwig in Ref.~\cite{Ludwig_RelatingNogCtoPosC}, we can recast the $c=-2$ electronic operators into our notations:
\begin{equation} \label{eq:GuruLudwig_Elec}
\begin{split}
\mathcal{V}_{-\lambda}^{\uparrow, \rm GL} & = \sum_{n} \sqrt{|n|} \psi_{n}^\dagger \cdot \mathbf{V}_{-n-\lambda}^c \, , \\  \mathcal{V}_{-\lambda}^{\downarrow, \rm GL} & = \sum_{n} \sign (-n) \sqrt{|n|} \, \psi_n \cdot  \mathbf{V}_{-n-\lambda}^c \, .
\end{split}
\end{equation} Compared to $\mathcal{V}_{-\lambda}^{\uparrow}$ and $\mathcal{V}_{-\lambda}^{\downarrow}$, the contribution of the derivative has been spread in a more symmetric way among the two spin species (Note that we could also split the $\sign(-n)$ at the price of dealing with complex numbers). The operators $\mathcal{V}_{-\lambda}^{\uparrow, \rm GL}$ and $\mathcal{V}_{-\lambda}^{\downarrow, \rm GL}$, or combinations of them, are also used in the free fermionic models of Refs.~\cite{LeeWen_HaldaneRezayiWithoutCFT,MilovanovicRead_EdgePairedStates}. While the SU(2)-symmetry of the underlying microscopic model is more obvious in this formalism, the electronic operators become non-local objects. These long distance behaviors are interpreted as indicators of the HR phase criticality~\cite{ReadGreen_PariedFermions}, although no rigorous proof or convincing numerical evidence have been able to show the bulk gaplessness yet. Once turned into an MPS, we numerically found the exact same results with either of the two representations Eq.~\ref{eq:ModeExpansion_ElecOperators} and Eq.~\ref{eq:GuruLudwig_Elec}. Nevertheless, all demanding computations were only performed with the prescription of Sec.~\ref{ssec:OurCFTDescription}, \textit{i.e.} with the local electronic operators $\mathcal{V}^\uparrow$ and $\mathcal{V}^\downarrow$.

\subsection{SU(2) Invariance}

We now investigate the spin-singlet nature of the HR state within our formalism. First note that Eq.~\ref{eq:HaldaneRezayiFirstQuantized} only describes a system of indistinguishable fermions after antisymmetrization over both the electronic spin and position. This procedure is accounted for in the CFT language by the commutation relation and OPE between electronic operators, as shown in Ref.~\cite{OurPaper_HalperinMPS}. Defining the electronic spinor $\mathcal{W} (w_i) = \mathcal{V}^\uparrow(w_i) \ket{\uparrow_i} +  \mathcal{V}^\downarrow(w_i) \ket{\downarrow_i} $ with $\ket{\uparrow_i}$ and $\ket{\downarrow_i}$ the two spin states of the $i$-th particle, the fully antisymmetric HR WF reads:
\begin{equation} \label{eq:HR_FirstQuant_AntisymSpin}
	\ket{\Psi^{\rm HR} (w_1, \cdots , w_{2N})} = \left\langle \mathcal{O}_{\rm Bkg} \prod_{i=1}^{2N} \mathcal{W}(w_i) \right\rangle \, .
\end{equation} 
Showing that Eq.~\ref{eq:HR_FirstQuant_AntisymSpin} describes a spin-singlet can be achieved as follows. We would like to find operators in the CFT whose actions on Eq.~\ref{eq:HR_FirstQuant_AntisymSpin} correspond to those of the total spin operators $S^z$ and $S^-$. They will allow for a direct evaluation of the quantities $\braOket{\Psi^{\rm HR}}{S^\sigma}{\Psi^{\rm HR}}$ with $\sigma \in \{z, +, - \}$, which should be zero for a spin-singlet. In App.~\ref{app:SU2SymmetryHR}, we exhibit such CFT counterparts of the total spin operators. These operators satisfy some Ward identities from which the equations $S^\sigma\ket{\Psi^{\rm HR}} =0$ are derived.

\section{MPS on the Infinite Cylinder} \label{sec:HR_Cylinder}

\subsection{Sketch of the Derivation}

We first briefly review the construction of exact MPS for spinful FQH model states written as CFT correlator. We refer the reader to Refs.~\cite{RegnaultEstienne_ExactMPS,OurPaper_HalperinMPS} for detailed derivations. In the Landau gauge, the cylinder LLL is spanned by the one-body states
\begin{equation} \label{eq:LLL_BasisCylinder}
	\phi_j (w) = \dfrac{e^{-(\gamma \ell_B)^2 j^2/2}}{\sqrt{L \ell_B \sqrt{\pi}}} e^{- \gamma j w} \, ,
\end{equation} which are labeled by $j \in \mathbb{Z}$ which fixes both the single particle momentum $k_j = \gamma j$ along the compact dimension and the orbital center on the cylinder axis $x_j = \gamma j \ell_B^2$ ($\ell_B$ is the magnetic length). Because they are plane wave $\phi_j (w) \propto e^{- \gamma j w}$, expanding all the electronic operators in the CFT correlator (see Eq.~\ref{eq:HR_FirstQuant_AntisymSpin}) into modes can be seen as inserting a $\mathcal{V}_{-\lambda}^\sigma$ operator for each orbital $\lambda$ occupied with a spin $\sigma$ electron. Denoting as $\ket{(n_k^\uparrow, n_k^\downarrow)_{k\in \mathbb{Z}} \rangle }$ the many-body occupation basis and reordering the various terms thanks to the electronic operator anticommutation relations, we get the site-dependent MPS form
\begin{equation}
\braket{\langle (n_k^\uparrow, n_k^\downarrow)_{k\in \mathbb{Z}}}{\Psi^{\rm HR} \rangle} = \left\langle \mathcal{O}_{\rm Bkg} \prod_{k \in \mathbb{Z}} A^{(n_k^\uparrow, n_k^\downarrow)} [k] \right\rangle \, ,
\end{equation} where the matrices read
\begin{equation} \label{eq:MPS_Cylinder_SiteDependent}
A^{(n^\uparrow, n^\downarrow)} [k] = \dfrac{1}{\sqrt{n^\uparrow ! n^\downarrow !}} \left( \dfrac{1}{\phi_k(0)} \mathcal{V}_{-k}^\uparrow \right)^{n^\uparrow} \left( \dfrac{1}{\phi_k(0)} \mathcal{V}_{-k}^\downarrow \right)^{n^\downarrow} \, .
\end{equation} 
The background charge for a system with $N_{\rm o}$ orbitals (\textit{i.e.} $N_{\rm o}-1$ flux quanta) is $\mathcal{O}_{\rm Bkg} = U_c^{N_{\rm o}}$. It can spread equally between orbitals using the relation $U_c \mathcal{V}_{-k}^\sigma = \mathcal{V}_{-(k-1)}^\sigma U_c$, $\sigma \in \{\uparrow, \downarrow\}$. The geometrical factors $\phi_k(0)$ in Eq.~\ref{eq:MPS_Cylinder_SiteDependent} can be accounted for by the insertion of
\begin{equation} \label{eq:U_Geometric_Cylinder}
	U_g = e^{-(\gamma \ell_B)^2 L_0} \, ,
\end{equation} between each orbitals~\cite{ZaletelMong_MPS,RegnaultEstienne_ExactMPS}, where $L_0$ is the zero-th Virasoro mode of the total CFT. Collecting the pieces, we obtain the orbital independent form
\begin{equation}
\braket{\langle (n_k^\uparrow, n_k^\downarrow)_{k\in \mathbb{Z}}}{\Psi^{\rm HR} \rangle} = \left\langle \prod_{k \in \mathbb{Z}} B^{(n_k^\uparrow, n_k^\downarrow)}  \right\rangle \, ,
\end{equation} with the following iMPS matrices
\begin{equation} \label{eq:MPS_Cylinder_SiteIndependent}
B^{(n^\uparrow, n^\downarrow)}  = \dfrac{1}{\sqrt{n^\uparrow ! n^\downarrow !}} \left( \mathcal{V}_0^\uparrow \right)^{n^\uparrow} \left(  \mathcal{V}_0^\downarrow \right)^{n^\downarrow} U_c U_g \, .
\end{equation}
The matrix elements of the iMPS matrices $B^{(n_k^\uparrow, n_k^\downarrow)}$ on the CFT basis given in Eq.~\ref{eq:CFT_Hilbert} can be evaluated analytically using the commutation relations of the bosonic and fermionic modes. However, the CFT Hilbert space is infinite and it must be truncated for any numerical simulations. The appended operator $U_g$ exponentially suppresses the contributions of highly excited CFT states. It thus seems natural to keep all states of conformal dimension no greater than a truncation parameter $P_{\rm max}$. The truncated iMPS matrices can be used to perform simulation on the infinite cylinder. At a fixed perimeter, we require the numerical convergence of the computed quantities with respect to $P_{\rm max}$. The truncation of the auxiliary space is constrained by the entanglement area law~\cite{Cirac_AreaLaw}, namely the bond dimension should grow exponentially with the cylinder perimeter $L$ to accurately describe the model WF (at least for a gapped bulk)~\cite{ZaletelMong_MPS}.

\subsection{The HR State is Gapless} \label{ssec:HR_gapless}

Using the MPS formulation of the HR state Eq.~\ref{eq:MPS_Cylinder_SiteIndependent}, we can now probe its gapless nature. We first detail the transfer matrix formalism, which allows to test thermodynamic properties of FQH systems~\cite{RegnaultEstienne_CorrelationLength}, and then numerically extract the bulk correlation lengths of the HR phase.

\subsubsection{Transfer Matrix Formalism}

A crucial object for iMPS calculation is the transfer matrix
\begin{equation}
	E = \sum_{n^\uparrow,n^\downarrow} B^{(n^\uparrow,n^\downarrow)} \otimes  \left(B^{(n^\uparrow,n^\downarrow)} \right)^* \, ,
\end{equation} where the complex conjugation is implicitly taken with respect to the CFT Hilbert space basis of Eq.~\ref{eq:CFT_Hilbert}. The transfer matrix is in general not Hermitian and might contain non-trivial Jordan blocks. It is however known that its largest eigenvalue in modulus is real and positive, and that the corresponding right and left eigenvectors can be chosen to be positive matrices~\cite{TransferMatrix_PositiveEigenvect}. The transfer matrix is particularly useful when computing expectation values of operators with finite support. We exemplify how such calculation is performed with the standard example of scalar products between MPS. Consider the MPS obtained for a finite number of orbitals $N_{\rm o}$ with boundary conditions  $(\alpha_L,\alpha_R)$ in the CFT Hilbert space:
\begin{equation}
\begin{split}
\ket{ \Phi_{\alpha_R}^{\alpha_L} \rangle } & = \sum c^{\alpha_R , \alpha_L} \ket{ (n_1^\uparrow,n_1^\downarrow) \cdots (n_{N_{\rm o}}^\uparrow,n_{N_{\rm o}}^\downarrow) \rangle } \, , \\ & c^{\alpha_R , \alpha_L}= \braOket{\alpha_L}{ B^{(n_1^\uparrow,n_1^\downarrow)} \cdots B^{(n_{N_{\rm o}}^\uparrow,n_{N_{\rm o}}^\downarrow)} }{\alpha_R} \, .
\end{split}
\end{equation} 
The overlap between any two of these MPS is given by
\begin{equation} \label{eq:OverlapTransferMatrix}
	\braket{ \langle \Phi_{\beta_R}^{\beta_L}}{ \Phi_{\alpha_R}^{\alpha_L} \rangle } = \braOket{\alpha_L, \beta_L^*}{E^{N_{\rm o}}}{\alpha_R,\beta_R^*} \, .
\end{equation}
In the limit of infinite cylinder $N_{\rm o} \to \infty$, the overlaps of Eq.~\ref{eq:OverlapTransferMatrix} are dominated by the largest eigenvectors of the transfer matrix. Note that the positivity of the largest eigenvector of $\mathcal{E}$ is coherent with its interpretation as an overlap matrix. Generically, most of the relevant physical information lies in the first leading eigenvalues and eigenvectors of the transfer matrix, making it a powerful numerical tool to extract physical properties of an infinite system.

As another example, consider a generic local operator $\mathcal{O}(x)$. At finite perimeter $L$, the MPS form obtained at truncation $P_{\rm max}$ necessarily leads to an exponential decay of its correlation function~\cite{Cirac_SimulabilityMPS}:
\begin{equation}
\langle \mathcal{O}(x)\mathcal{O}(0)\rangle - \langle \mathcal{O}(x) \rangle \langle \mathcal{O}(0) \rangle \propto e^{- |x| / \xi(L, P_{\rm max}) } \, .
\end{equation} 
The correlation length $\xi(L, P_{\rm max})$ is related to the ratio of the two largest eigenvalues, $\lambda_1(L, P_{\rm max})$ and $\lambda_2(L, P_{\rm max})$, of the transfer matrix~\cite{GapTransferMatrix}:
\begin{equation}
\xi(L, P_{\rm max}) = \dfrac{2 \pi \ell_B^2}{L \log \left| \frac{\lambda_1(L, P_{\rm max})}{\lambda_2(L, P_{\rm max})} \right|} \, .
\end{equation} It converges to a finite value in the thermodynamic limit, obtained for $P_{\rm max} \to \infty$ and $L/\ell_B \to \infty$ (in that order), for a gapped phase.

\subsubsection{Correlation Lengths}

\begin{figure*}
	\centering
	\includegraphics[width=0.95\textwidth]{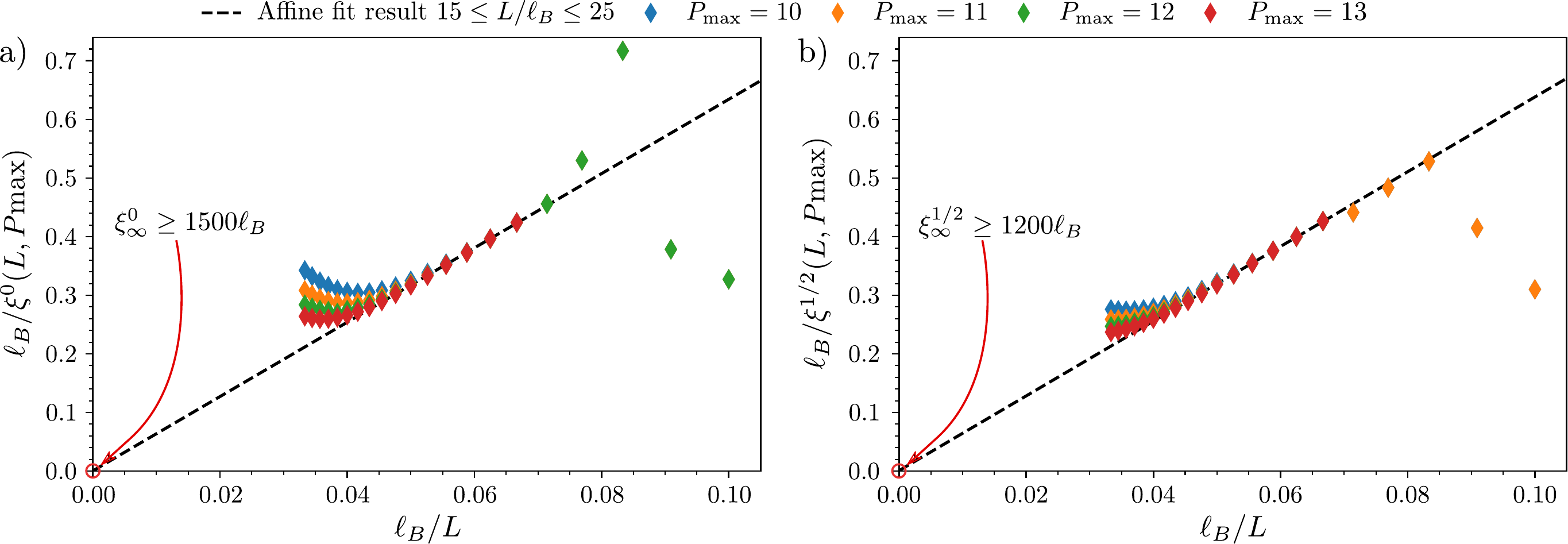}
	\caption{\emph{Inverse correlation length $1/\xi^a(L, P_{\rm max})$ in the charge sectors a) $a=0$ and b) $a=\sfrac{1}{2}$ as a function of the inverse cylinder perimeter $\ell_B / L$. To extract the thermodynamic values, we only keep the points that have converged to better than 2\% with respect to $P_{\rm max}$. At truncation parameter $P_{\rm max}=13$ it corresponds to $L/\ell_B \in [15,25]$. Affine and linear fit equally well capture our data, hinting toward an infinite bulk correlation of the HR state in the thermodynamic limit: $1/\xi_\infty^0 \to 0$ and $1/\xi_\infty^{\sfrac{1}{2}} \to 0$.} }
	\label{Fig:GaplessExcitations}
\end{figure*}

For topologically ordered phases of matter, the GS degeneracy leads to multiplicities in the transfer matrix eigenvalues, which can be resolved by splitting the CFT Hilbert space into topological sectors. Each of these sectors contains a single leading eigenvector of the transfer matrix and is stable under the action of the electronic operators. They are connected to each other by the deconfined anyonic excitations which leave the GS manifold stable~\cite{NayakWilczek_SpinorBraiding}. We can thus benefit from the structure of the CFT Hilbert space in numerical simulation. As discussed in Sec.~\ref{ssec:CFT_HilbertSpace}, the four charge sectors are stable under the action of the electronic operators and  $U_c$ shifts the U(1)-charge by one unit. It is therefore better suited for our calculation to consider the transfer matrix over two orbitals
\begin{equation}
E^2 = \sum_{\vec{n}^\uparrow,\vec{n}^\downarrow} \mathbf{B}^{(\vec{n}^\uparrow,\vec{n}^\downarrow)} \otimes  \left(\mathbf{B}^{(\vec{n}^\uparrow,\vec{n}^\downarrow)} \right)^* \, ,
\end{equation} with $\mathbf{B}^{(\vec{n}^\uparrow,\vec{n}^\downarrow)} = B^{(m_{1}^\uparrow,m_{1}^\downarrow)} B^{(m_{2}^\uparrow,m_{2}^\downarrow)}$. The bold indices stand for the occupation numbers of two consecutive sites: $\vec{n}^\uparrow = (m_{1}^\uparrow,m_{2}^\uparrow)$ and $\vec{n}^\downarrow = (m_{1}^\downarrow,m_{2}^\downarrow)$. The $\mathbf{B}$ matrices are block diagonal with respect to the four charge sectors, giving to $E^2$ a similar block structure. We can thus target a specific block during the diagonalization of the transfer matrix, improving the numerical efficiency.

We observe that $E^2$ has eight degenerate leading eigenvectors, two in each charge sector $a\in \{0,1/2,1,3/2\}$. The two-fold degeneracy in each sector can be further resolved by focusing on the hierarchy-like topological sectors the the HR phase (see Sec.~\ref{ssec:CFT_HilbertSpace}). Related by a center of mass translation on the cylinder or by a spin symmetry~\cite{OurPaper_HalperinMPS}, the four sectors in $a=0$ and $a=1$ (resp. $a=1/2$ and $a=3/2$) share the same correlation length that we denote as  $\xi^0$ (resp. $\xi^{\sfrac{1}{2}}$). We have numerically extracted these correlation lengths as a function of the truncation parameter $P_{\rm max}$ and the cylinder perimeter $L$. The results depicted in Fig.~\ref{Fig:GaplessExcitations} show that the different correlation lengths grow linearly with the cylinder perimeter. The thermodynamic values are extracted by affine extrapolation at $\ell_B/L \to 0$ (over the points that have converged with respect to the bond dimension, \textit{i.e.} $L \leq 25 \ell_B$). We find $1/\xi_\infty^a \simeq 0$ in all sectors, and observe that affine and linear functions fit equally well our data. This diverging correlation length of the HR state in the thermodynamic limit reveals its gapless nature in all the sectors. Such a feature prevents the HR to describe a quantized Hall plateau at half filling of a given Landau level. Still, the HR state could remain relevant at a two-dimensional critical point such as the weak to strong $d$-wave phase transition~\cite{ReadGreen_PariedFermions}.

We would like to make a few remarks. First, our results show that, although it stabilizes the gapped Laughlin phase in the spin polarized case~\cite{Kivelson_PseudoPotentials} and despite its physical relevance~\cite{HaldaneRezayi_Wavefunction}, the SU(2)-symmetric hollow-core model Hamiltonian is gapless. Microscopically, this non-trivial result hints that the contact interaction between electrons with opposite spins is necessary to make the model FQH state incompressible. The addition of a $V_0$ pseudopotential was considered numerically~\cite{Jain_SpinSinglet,Regnault_dWavePairing}, and shown to energetically favor Jain's spin-singlet state when $V_0 \sim V_1$. Finally, we remark that the transfer matrix only has eight degenerate leading eigenvectors and not ten, as would be expected from the HR GS degeneracy on the torus. We will elaborate on this issue in Sec.~\ref{ssec:TenFold_Degenerate}, but already state that the missing information is contained in a Jordan block which is not resolved during the numerical diagonalization of the transfer matrix.

\subsection{Entanglement Entropy} \label{ssec:iMPS_EE}

\begin{figure}
	\centering
	\includegraphics[width=\columnwidth]{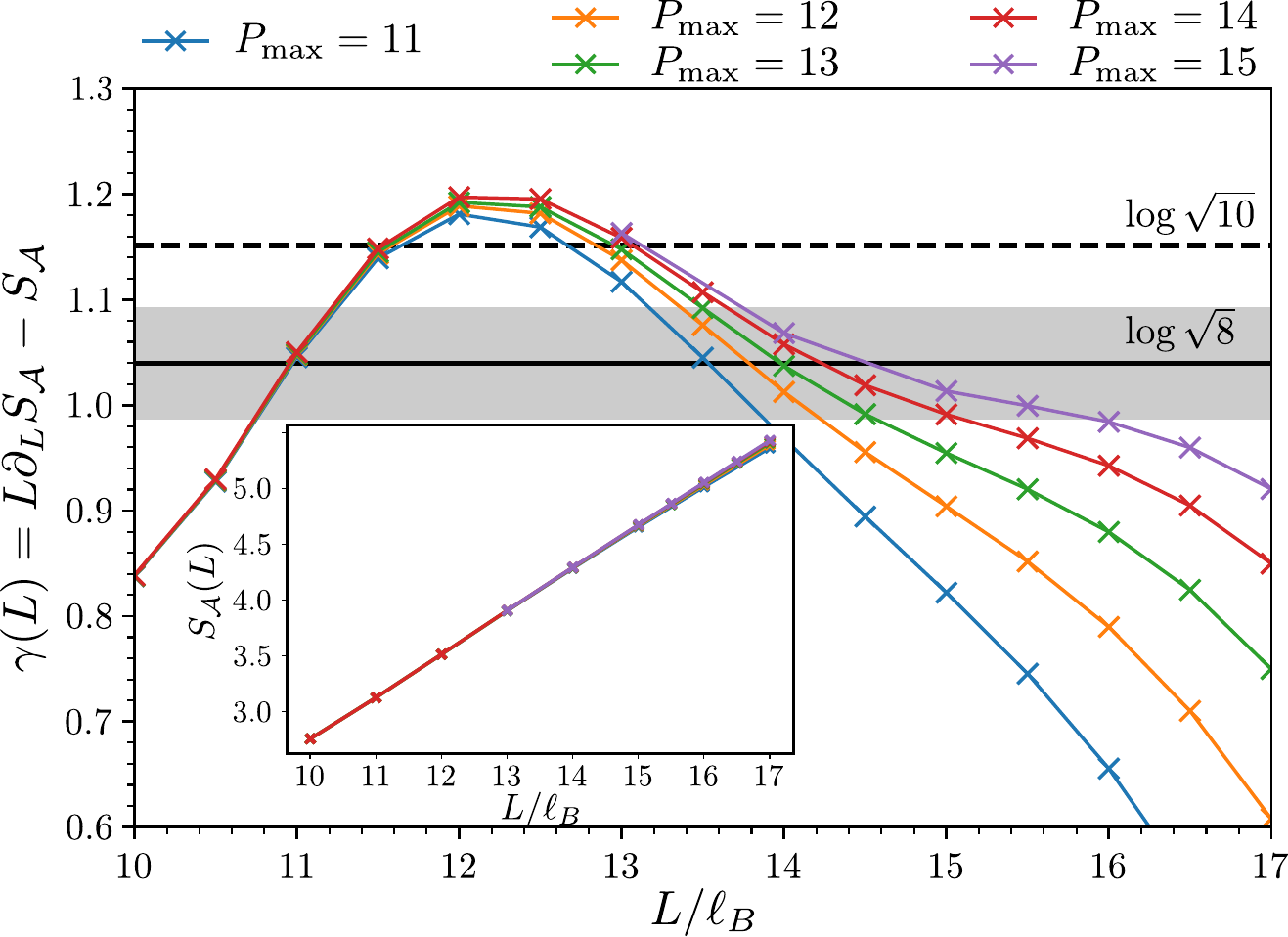}
	\caption{\emph{Numerical extraction of the entanglement entropy $S_{\mathcal{A}}$ for a half infinite cylinder in the topological sector $a=1$ with even fermionic parity. Main: Correction to the area law $\gamma (L)$, numerically extracted with finite differences, as a function of the cylinder perimeter $L$. Due to the critical nature of the HR state, the results are plagued with finite size effects and we often cannot reach convergence as a function of the truncation parameter $P_{\rm max}$. However, the numerically extracted points seem to oscillate around the value $\log \sqrt{8}$ (solid line, gray shaded area shows a five percent uncertainty around this value) and are hardly consistent with $\log \sqrt{10}$ (dashed line). They  Inset: No deviation of the entanglement entropy $S_{\mathcal{A}}$ from the area law are detected within the considered range of perimeters. }}
	\label{Fig:EntanglementEntropy_iMPS}
\end{figure}

Discussing about adiabatic braiding of excitations in the HR phase might not be meaningful because of its gapless nature. Consequently, statements about the underlying topological order or the universality of long range entanglement in the HR phase should be done with caution. However, iMPS calculation set a natural cut-off through the finite perimeter $L$. It is thus relevant to investigate the consequences of criticality for the eight GS that we have obtained on the infinite cylinder thank to entropic measurements. Because the correlation length is proportional to the cylinder perimeter (see Fig.~\ref{Fig:GaplessExcitations}), our numerical results are plagued with large finite size effects, making it difficult to extract thermodynamic features of the HR phase.

We exemplify our study on the GS obtained on the infinite cylinder in the topological sector $a=1$ with even fermionic parity (see Sec.~\ref{ssec:CFT_HilbertSpace}). We consider a bipartition of the cylinder $\mathcal{A}-\mathcal{B}$ into two halves, with $\mathcal{A} = \{ w=x+iy \, | \, x <0 \, , y \in [0,L] \}$, and compute the corresponding Real-Space Entanglement Entropy (RSEE)~\cite{RSESdubail,RSESsterdyniak} $S_{\mathcal{A}}(L)$ with the techniques developed in Ref.~\cite{OurNatComms_H3L3}. For a topologically ordered fully-gapped bulk GS, it follows an area law
\begin{equation} \label{eq:AreaLaw}
	S_{\mathcal{A}}(L) = \alpha L - \gamma \, ,
\end{equation}
where $\alpha$ is a non-universal parameters and $\gamma$, the Topological Entanglement Entropy (TEE), characterizes the topological order~\cite{TopoCorrectionKitaev,TopoCorrectionLevinWen}. With the reachable system sizes, we have not detected any deviations to Eq.~\ref{eq:AreaLaw} (see the inset of Fig.~\ref{Fig:EntanglementEntropy_iMPS}). We also extracted the first correction to the area law with finite differences as $\gamma (L) = L\partial_L S_{\mathcal{A}} - S_{\mathcal{A}}$, the results are displayed in Fig.~\ref{Fig:EntanglementEntropy_iMPS}. As stated above, the strong finite size effects impose to consider large perimeters where convergence with respect to the truncation $P_{\rm max}$ is hard to reach, especially for subleading quantities such as $\gamma(L)$. The results for $L\geq 14\ell_B$ however seem to show that $\gamma(L)$ reaches a plateau around $\log\sqrt{8}$ when $L$ increases. Using a slightly different extrapolation method which filters out the small system sizes, the presence of the plateau at $\log\sqrt{8}$ is even more convincing as shown in App.~\ref{app:AdditionalResults}. This is the expected theoretical value for a topological phase governed by the \textbf{K}-matrix given by Eq.~\ref{eq:Kmatrix_331}. Although we can not rule out the possibility of unnoticed logarithmic corrections to the RSEE, these results are reminiscent of those of the Gaffnian state~\cite{Simon_Gaffnian}. Both states are non-Abelian and built from non-unitary CFT. In both cases, the TEE seems to only capture the Abelian part of the phase~\cite{RegnaultEstienne_CorrelationLength}. 

As a last remark, we note that the two topological sectors arising from the Jordan block structure (see Sec.~\ref{ssec:TenFold_Degenerate}) do not seem to contribute to the total quantum dimension $\mathcal{D}$ of the phase. Indeed, we would have $\mathcal{D}\geq 10$ if they did, with equality if all sectors were Abelian. Our results for any of the eight hierarchy-like states displayed in Fig.~\ref{Fig:EntanglementEntropy_iMPS} are not consistent with values above $\log\sqrt{10} \simeq 1.15$.

\section{CFT Model States on the Torus} \label{sec:MPS_Torus}

For pedagogical purposes, we focus in this section on spinless fermionic systems to illustrate the construction of exact MPS for FQH model state on the torus geometry. The results derived can be straightforwardly extended to spinful and/or bosonic systems. 

\subsection{Particles in a Magnetic Field} 

\subsubsection{Boundary Conditions on Torus}

We first consider the problem of a single particle in a magnetic field, using the Landau gauge $\vec{A} = \ell_B^{-2} (0,x)$. The particle is free to move on the torus, with Hamiltonian $H_{\rm L} = ( \vec{p} - \vec{A} )^2/2$, which imposes some constraints on the dynamics that we now derive. The torus is mathematically obtained as the quotient of the complex plane by a two-dimensional lattice generated by $\vec{L}_1 = ( 0 , L_1 )$ and $\vec{L}_2 = ( - L_2 \sin\theta , L_2 \cos\theta )$: 
\begin{equation}
\mathbb{T}^2 = \mathbb{C} / \left( i L_1 \mathbb{Z} + i e^{i\theta} L_2 \mathbb{Z} \right) \, .
\end{equation} That is, we work on the complex plane and identify $w = w + i L_1 = w + i e^{i\theta} L_2$. The unusual factors '$i$' are rather conventional and are included for consistency with the cylinder $\mathbb{C}/iL_1 \mathbb{Z}$. The torus is characterized by its aspect ratio
\begin{equation}
\tau = \dfrac{L_2}{L_1} e^{i\theta} \, , \quad \Im \tau > 0 \, .
\end{equation}  The torus geometry imposes the constraints $|\phi (w)| = |\phi (w + i L_1)| = |\phi (w + i e^{i\theta} L_2)|$ on any torus one-body WF $\phi$. The equation only involves the magnitude of $\phi$ because the different points of the quotient lattice are related by non-trivial gauge transformations~\cite{HaldaneRezayi_TorusGeom}. This may be understood considering the translation operator by $\vec{l}$ in presence of a magnetic field
\begin{equation}
t\left( \vec{l} \right) = \exp\left[ \vec{l} \cdot \left( \nabla - i \vec{A} \right) - i \ell_B^{-2} \, \vec{l} \times \vec{r} \right] \, ,
\end{equation} $\vec{a} \times \vec{b} = a_1 b_2 - a_2 b_1$ denotes the cross products of the vectors $\vec{a}=(a_1,a_2)$ and $\vec{b}=(b_1,b_2)$. We assume that no net fluxes pass through the torus' non-contractible loops, such that the Torus Boundary Conditions (TBC) are $t(\vec{L}_1) \ket{\phi \rangle} = t(\vec{L}_2) \ket{\phi \rangle} = \ket{\phi \rangle}$~\cite{ReadRezayi_ZeroModes}. Evaluating these equations at position $w = x + i y$ gives:
\begin{align} \label{eq:MagneticTranslation_BC}
& \phi (w + i L_1) = \phi (w) \, , \\ 
& \phi (w + i e^{i\theta} L_2) =  \exp\left[ -i \dfrac{L_1 L_2 \sin\theta}{\ell_B^2} \left( \frac{\Re\tau}{2} + \frac{y}{L_1} \right) \right] \phi (w) \, . \notag
\end{align} These quasi-periodic boundary conditions simply transcribes that $\vec{A}$ cannot be globally defined on $\mathbb{T}^2$, as this would lead to $\int_{\mathbb{T}^2} B \, {\rm d}^2 w = 0$. In a more geometric language, the WF is a section of a non-trivial line bundle over the torus.

\subsubsection{Discrete Magnetic Translations}

Consistency of the TBC implies restrictions on the magnetic field and on the physically allowed magnetic translations. Contrary to the plane geometry, infinitesimal translations are not consistent with the TBC of the WF. They change the physical properties of the system by adding fluxes through the torus' non-contractible loops. It is well known that consistency with respect to the TBC leads to a \emph{discrete} set of physically acceptable magnetic translation operators.

Magnetic translations satisfy the Girvin-MacDonald-Platzman algebra~\cite{GMP_Algebra,RegnaultBernevig_TorusSymmetries}:
\begin{equation} \label{eq:CommutationRelation_MagneticOpe}
\left[ t\left( \vec{l}_1 \right) , t\left( \vec{l}_2 \right) \right] = 2i \sin\left(\dfrac{\vec{l_1} \times \vec{l_2}}{2 \ell_B^2}\right)  t\left( \vec{l}_1 + \vec{l}_2 \right) \, .
\end{equation} 
Going around the torus' principal region should give the identity, requiring that $t\left( \vec{L}_1 \right)$ and $t\left( \vec{L}_2 \right)$ commute. Hence, using Eq.~\ref{eq:CommutationRelation_MagneticOpe}, the magnetic flux threading the torus should be a multiple integer $N_\phi \in \mathbb{N}$ of the flux quantum, \textit{i.e.} 
\begin{equation}
N_\phi = \dfrac{\left|\vec{L}_1 \times \vec{L}_2\right|}{2\pi \ell_B^2} \, .
\end{equation} 
Similarly, the physically allowed magnetic translations preserve the TBC and should commute with $t\left( \vec{L}_1 \right)$ and $t\left( \vec{L}_2 \right)$. This discrete set of allowed magnetic translations can be obtained from Eq.~\ref{eq:CommutationRelation_MagneticOpe} and are generated by the two translations:
\begin{equation} \label{eq:MagneticTranslation_Allowed}
	t_1 = t\left( \vec{L}_1/N_\phi \right) \, , \quad t_2 = t\left( \vec{L}_2/N_\phi \right)  \, , 
\end{equation} which satisfy $t_1 t_2 = \exp(2i \pi/N_\phi) t_2 t_1$.

\subsubsection{Landau Problem}

The Landau problem on the torus still retains the usual harmonic oscillator structure~\cite{Laughlin_QuantizedHallConductance}. In particular with our gauge choice, we have
\begin{equation}
	H_{\rm L} = \hbar \omega_c \left( a^\dagger a + \frac{1}{2} \right) \, , \quad a = \frac{1}{\sqrt{2}\ell_B} \left( 2 \ell_B^2 \, \overline{\partial} + x \right) \, .
\end{equation} 
Sending the cyclotron energy $\hbar \omega_c$ to infinity projects the system to the LLL. The latter consists of all the states $\phi$ which are annihilated by $a$ and obey the proper TBC. They are of the form:
\begin{equation} \label{eq:LLL_Torus_GeomFactors}
\phi (w) = \exp\left( - \frac{x^2}{2 \ell_B^2} \right) f(w) \, , 
\end{equation} where $f(w)$ is a holomorphic function satisfying the boundary conditions $f(w + i L_2 e^{i\theta}) = e^{-2i\pi N_\phi (\frac{w}{i L_1} + \frac{\tau}{2}) } f(w)$ and $f(w + i L_1) = f(w)$ as inferred from Eq.~\ref{eq:MagneticTranslation_BC}. In App.~\ref{app:HolomorphicLLL}, we show that the LLL hosts $N_\phi$ orbitals. Because $t_1$ commutes with $H_{\rm L}$, we choose a LLL basis made of $t_1$ eigenvectors:
\begin{equation} \label{eq:LLL_Torus_t1}
f_{k_y} (w) = \dfrac{1}{\sqrt{L_1 \ell_B \sqrt{\pi}}} \vartheta \begin{bmatrix} k_y/N_\phi \\	0 \end{bmatrix} \left( \frac{N_\phi}{i L_1} w \Big| N_\phi \tau \right) \, ,
\end{equation} with $k_y =  0 , \cdots , N_\phi -1 $ and where $t_1 f_{k_y} = \exp(2i\pi k_y/N_\phi) f_{k_y}$ comes out of the properties of the Jacobi theta function $\vartheta$. The other primitive translation acts as $t_2 f_{k_y} = f_{k_y+1}$. Amongst other things, this implies that the chosen LLL orbitals share the same norm, which is of importance for the expression of the MPS (see Sec.~\ref{ssec:MPS_Torus_Demo}). Expanding $f_{k_y}$ gives a intuitive understanding of Eq.~\ref{eq:LLL_Torus_t1} as the periodic counterpart of the cylinder orbitals (compare with Eq.~\ref{eq:LLL_BasisCylinder}):
\begin{equation}
f_{k_y} (w) = \dfrac{1}{\sqrt{L_1 \ell_B \sqrt{\pi}}} \sum_{j \in k_y + N_\phi \mathbb{Z}} e^{i\pi\tau \frac{j^2}{N_\phi}} e^{\frac{2\pi}{L_1} j w} \, .
\end{equation}

\subsection{Model WFs as Conformal Blocks}

In the last paragraph, we saw that the LLL enjoys a holomorphic structure which, in the Landau gauge, has strict periodic conditions in the $\vec{L}_1$ direction (see Eq.~\ref{eq:MagneticTranslation_BC}). We can thus picture the torus as a finite cylinder of perimeter $L_1$ whose ends have been glued together with a twist which depend on $\tau$~\cite{YellowBook,PollmannZaletel_TopoCharacTorus}. We will thus continue to use CFTs defined on the cylinder, as in Sec.~\ref{sec:HR_Construction}, and impose that the physical WFs satisfy the TBC. We recall for that purpose the conformal mapping from the cylinder coordinate $w= x+iy$ to the plane $z = \exp\left(\frac{2 \pi}{L_1} w \right)$. The translation $w \to w + i L_2 e^{i \theta}$ becomes a rotation dilation $z \to qz$ with $q=e^{2i\pi \tau}$.

We now consider a system of $N_e$ fermions and $N_\phi$ flux quanta, thus at filling fraction $\nu = N_e/N_\phi$. We focus on model FQH WFs whose underlying CFT separates the neutral and charge degrees of freedoms~\cite{RegnaultEstienne_ExactMPS}. The electronic operator generically reads $\mathcal{V} = \Psi \cdot : e^{i R \varphi} :$ where $\Psi$ only acts on the neutral CFT, and $\varphi$ is a chiral massless bosonic field with compactification radius $R=\nu^{-1/2}$. We assume that the electronic operator at different positions anticommute as we are interested in fermionic systems in this article. The results can be readily extended to bosonic and/or spinful cases. The model WF in a given topological sector $a$ on the torus takes the form:
\begin{equation} \label{eq:Torus_ConformalBlocks}
\Psi_a (w_1, \cdots , w_{N_e}) = \Tr_a \left[X \mathcal{V}(w_1) \cdots \mathcal{V}(w_{N_e}) \right] \, ,
\end{equation} with $ X = X_{\rm AR} X_{\rm P} X_{\rm Bkg}$ and where $\Tr_a (\cdots) = \Tr (P_a \cdots )$ denotes the trace in sector $a$ ($P_a$ being the projector on topological sector $a$). It assumes prior knowledge of the different existing topological sectors, and numerical simulations furthermore require a way to delineate the sectors within the chosen computational CFT basis in order to represent $P_a$. The operators $X_{\rm Bkg}$, $X_{\rm AR}$ respectively account for the charge neutrality in the CFT correlator, the anti-commutation relation of the fermions while their interplay with $X_{\rm P}$ produce the phase factors arising from the TBC. They read: 
\begin{subequations}\label{eq:X_OperatorsEndTorus} \begin{align}
		X_{\rm Bkg} &= e^{- i \sqrt{\nu} N_\phi \varphi_0} \, , \\
		X_{\rm AR} &= e^{- i \pi (N_e-1) \sqrt{\nu} a_0} \, , \\
		X_{\rm P} &= q^{L_0 + \frac{N_\phi}{2}\sqrt{\nu} a_0} \, , 
\end{align} \end{subequations} with $q=\exp(2i\pi\tau)$. We show in App.~\ref{app:TBCforConfBlocks} that that the many-body WFs of Eq.~\ref{eq:Torus_ConformalBlocks} indeed satisfy the TBC.

\subsection{MPS on the Torus}\label{ssec:MPS_Torus_Demo}

The MPS representation of Eq.~\ref{eq:Torus_ConformalBlocks} follows from expanding all the electronic operators into modes. Thank to Eq.~\ref{eq:MassageModeExpansionForMPS} derived in App.~\ref{app:TBCforConfBlocks}, we can rearrange the different sums into:
\begin{equation}
\begin{split}
& \Psi_a (w_1 ,\cdots ,w_{N_e}) = \\ 
& \sum_{s_i = 0}^{N_\phi-1} \Tr_a \left[X \mathcal{V}_{-s_1} \cdots  \mathcal{V}_{-s_{N_e}} \right] \prod_{i=1}^{N_e} e^{-i \pi \tau \frac{s_i^2}{N_\phi}} f_{s_i} (w_i) \, .
\end{split}
\end{equation} 
The electronic operator anticommutation relation allows us to order the $(s_i)_{i}$ to get
\begin{align}
\Psi_a (w_1 ,\cdots ,w_{N_e})  = \!\!\!\!\!\! & \sum_{N_\phi > s_1 > \cdots > s_{N_e} \geq 0} \!\!\!\!\!\! \Tr_a \left[X \prod_{i=1}^{N_e} e^{-i \pi \tau \frac{s_i^2}{N_\phi}} \mathcal{V}_{-s_i} \right] \notag \\
& \left( \sum_{\sigma \in \mathfrak{S}_e} \varepsilon(\sigma) \prod_{i=1}^{N_e}  f_{s_{\sigma(i)}} (w_i)  \right) \, ,
\end{align} where we have used partitions $\sigma \in \mathfrak{S}_e$ to treat all possible orderings and denoted as $\varepsilon(\sigma)$ their signature. The fully antisymmetric product of lowest LL WFs is the first quantized form of the many-body occupation basis
\begin{equation}
	\ket{m_0 , \cdots , m_{N_\phi-1} \rangle } = c_{s_1}^\dagger \cdots c_{s_{N_e}}^\dagger \ket{\Omega\rangle} \, , 
\end{equation} with $N_\phi > s_1 > \cdots > s_{N_e} \geq 0$ and where $c_{s}^\dagger$ creates a particles on orbital $s$ above the Fock vacuum $\ket{\Omega\rangle}$. We thus have a site-dependent MPS form for the model WFs on the torus:
\begin{align}
& \braket{\langle m_0 , \cdots , m_{N_\phi-1} }{\Psi_a \rangle} = \Tr_a \left[X \prod_{j=0}^{N_\phi -1} A_{\mathbb{T}}^{(m_j)}[j] \right] \, , \notag \\
& A_{\mathbb{T}}^{(m)}[j] = \dfrac{1}{\sqrt{m!}} \left( e^{-i \pi \tau \frac{j^2}{N_\phi}} \mathcal{V}_{-j} \right)^m \, .
\end{align}
As we previously did on the cylinder, we can turn this MPS into a site independent formulation by spreading $X_{\rm P} X_{\rm Bkg}$ equally between orbitals. More precisely, we finally reach 
\begin{align}\label{eq:MPS_Torus_SiteIndep}
& \braket{\langle m_0 , \cdots , m_{N_\phi-1} }{\Psi_a \rangle} = \Tr_a \left[X_{\rm AR} \prod_{j=0}^{N_\phi -1} B_{\mathbb{T}}^{(m_j)} \right] \, , \notag \\
& B_{\mathbb{T}}^{(m)} = \dfrac{1}{\sqrt{m!}} U_{\mathbb{T}} \left( \mathcal{V}_0 \right)^m \, , \quad U_{\mathbb{T}} = q^{\frac{L_0}{N_\phi}} e^{-i\sqrt{\nu}\varphi_0} \, .
\end{align}
As can be seen, the expressions of both $B_{\mathbb{T}}$ and $U_{\mathbb{T}}$ are similar to their counterparts on the cylinder.

\section{HR on the Torus: Zero Modes and Degeneracy} \label{sec:HR_10fold}

\subsection{Exact Diagonalizations}

\begin{figure}
	\centering
	\includegraphics[width=\columnwidth]{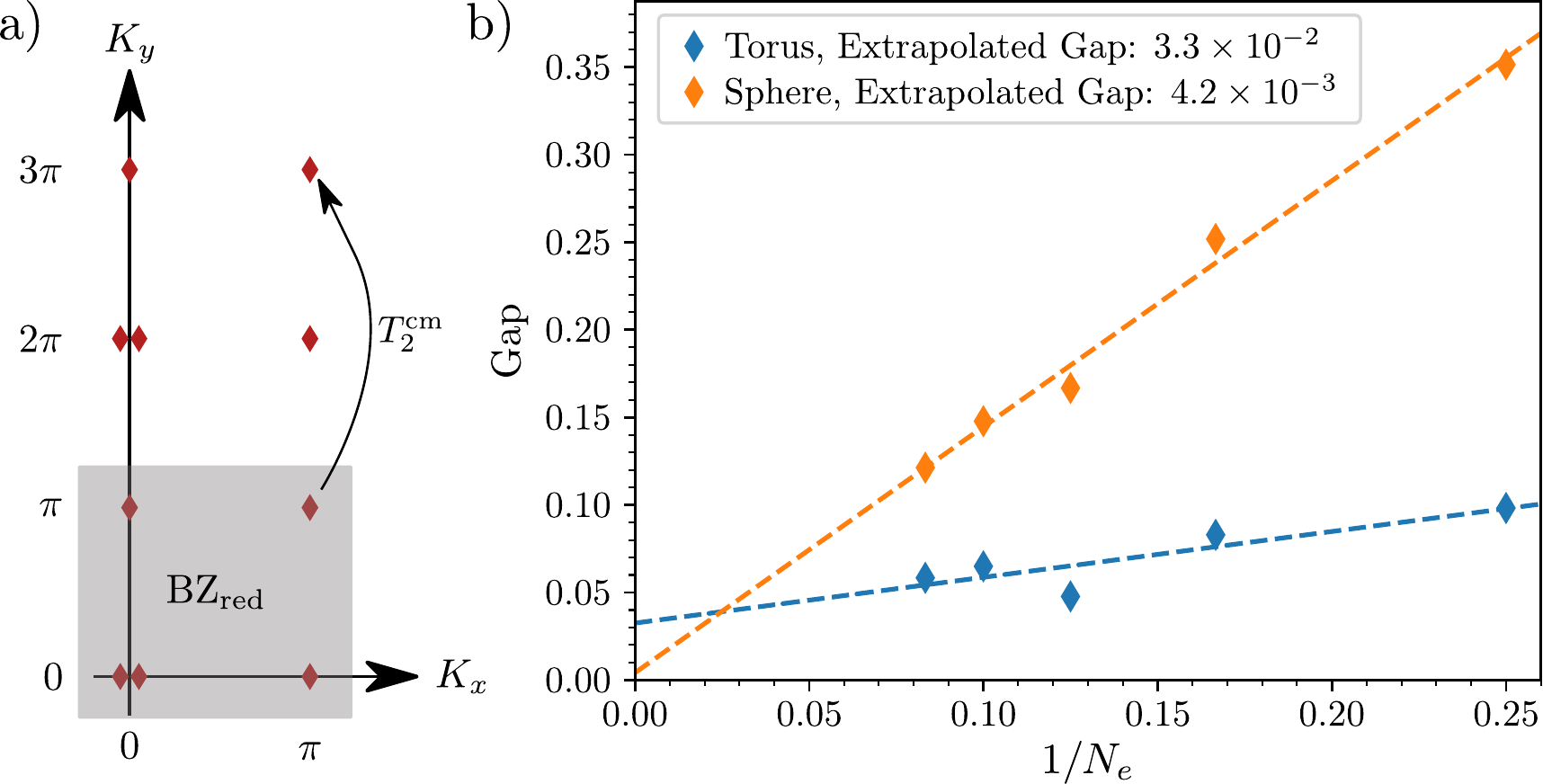}
	\caption{a) \emph{Structure of the HR ten-fold degenerate GS manifold on the torus. $(K_x, K_y)$ indicate the many-body momenta along the two directions of the torus. With center of mass translations, all the different GS can be obtained from the five states lying in the reduced Brillouin zone ${\rm BZ}_{\rm red}$ (gray shaded region).} b) \emph{Scaling of the gap on the torus (blue) and sphere (orange) geometries as a function of the inverse particle number. The dashed lines are linear fits to the data whose intercepts provide extrapolation of the gap in the thermodynamic limit $N_e \to \infty$. While the gap seems to close in the thermodynamic limit on the sphere, the reachable systems size are too small to see any strong signatures of the gapless nature of the HR phase in ED on the torus. } }
	\label{Fig:ED_TorusStudy}
\end{figure}

The system consists of $N_e=2N$ spin-\sfrac{1}{2} fermions on a torus pierced with $N_\phi =4N$ flux quanta, thus at filling fraction $\nu = \frac{1}{2}$. They occupy the lowest LL spanned by Eq.~\ref{eq:LLL_Torus_t1} and interact, irrespective of their spin, through a $V_1$ two-body pseudopotential (see Eq.~\ref{eq:V1_PseudoPot}). Many-body translation operators on the torus factorize into the product of relative and center-of-mass translations~\cite{Haldane_ManyBodyTransSym}. The latter are generated by \begin{equation}
T_1^{\rm cm} = \prod_{i=1}^{N_e} t_1^i \quad \text{and } \quad T_2^{\rm cm} = \prod_{i=1}^{N_e} t_2^i \, .
\end{equation} At filling factor $\nu=\frac{1}{2}$, $T_1^{\rm cm}$ and $(T_2^{\rm cm})^2$ commute with one another and with the hollow-core Hamiltonian~\cite{Haldane_ManyBodyTransSym,RegnaultBernevig_TorusSymmetries}. These many-body conservation laws make ED studies more efficient and allow to reach large system sizes. For the sake of clarity, we now focus on a rectangular torus ($\theta = \pi/2$), although the construction of Sec.~\ref{sec:MPS_Torus} applies to any other aspect ratio. The many-body eigenstates $\ket{\Psi (\vec{K}) \rangle}$ carry the associated momentum quantum number $\vec{K}$ and satisfy 
\begin{subequations} \begin{align}
T_1^{\rm cm} \ket{\Psi (\vec{K}) \rangle} = e^{i K_y / 2} \ket{\Psi (\vec{K}) \rangle} \\
(T_2^{\rm cm})^2 \ket{\Psi (\vec{K}) \rangle} = e^{i K_x} \ket{\Psi (\vec{K}) \rangle}
\end{align} \end{subequations} 
where the momentum $\vec{K}$ belongs to the Brillouin zone~\cite{RegnaultBernevig_TorusSymmetries}
\begin{equation}
\begin{split}
\left\{ \vec{K} = \left( \dfrac{K_x}{L_2} , \dfrac{K_y}{L_1} \right) \, \Big| \right. & \,  K_x = 0 , \cdots , (N_e -1) \dfrac{2\pi}{N_e} \, ;  \\
& \, \left. K_y = 0 , \cdots ,  (N_e -1) \dfrac{4\pi}{N_e}  \right\}  \,  . 
\end{split}
\end{equation}
Because $T_1^{\rm cm}$ and $T_2^{\rm cm}$ anticommute, $T_2^{\rm cm}$ relates an eigenstate at any eigenstate at $(K_x, K_y)$ to an eigenstate at $(K_x, K_y + 2 \pi)$ with the same energy. We can thus restrict our study to the reduced Brillouin zone
\begin{equation} \label{eq:ReducedBrillouinZone}
\begin{split}
{\rm BZ}_{\rm red} = & \left\{ \vec{K}  = \left( \dfrac{K_x}{L_2} , \dfrac{K_y}{L_1} \right)  \, \Big| \right. \\
& \qquad \left. K_x, K_y = 0 , \frac{2\pi}{N_e}, \cdots , (N_e -1) \frac{2\pi}{N_e}  \right\}  \, ,
\end{split}
\end{equation} depicted in Fig.~\ref{Fig:ED_TorusStudy}a.

The ED of the hollow core Hamiltonian shows that it has five zero energy states in ${\rm BZ}_{\rm red}$, and thus ten zero energy states on the torus. As depicted in Fig.~\ref{Fig:ED_TorusStudy}a, three of them are located respectively at $(\pi,0)$, $(0,\pi)$ and $(\pi,\pi)$ while the momentum $(0,0)$ hosts two degenerate zero energy states. We also considered the neutral gap of HR phase for the model interaction, as shown in Fig.~\ref{Fig:ED_TorusStudy}b for a square torus ($\tau=i$) and on the sphere. We have considered systems of up to $N_e=12$ particles on both geometries. While the gap on the sphere seems to converge to zero in the thermodynamic limit, in agreement with the MPS results of Sec.~\ref{ssec:HR_gapless}, its closing on the torus is not so clear. This apparent lack of gap closing is most probably due to the few reachable system sizes rather than an actual feature. Indeed, there is no reasons why the gapless excitations on the sphere should disappear on the torus. In both cases the MPS transfer matrix is essentially the same, comparing Eq.~\ref{eq:MPS_Cylinder_SiteIndependent} to Eq.~\ref{eq:MPS_Torus_SiteIndep}. Hence, we expect them to host similar low-lying excitations, which are within the single mode approximation tightly connected to the excited states of the transfer matrix.

In the following, we show that our construction accurately captures the whole HR physics and that the ten zero energy states of the model Hamiltonian can be written in our MPS formalism. There are a few obstacles that we should overcome. We shall first relate the many-body momentum $\vec{K}$ and the parameters of our MPS ansatz. This allows to reproduce the four hierarchy-like ground states of ${\rm BZ}_{\rm red}$. The last zero energy state in ${\rm BZ}_{\rm red}$ requires a careful treatment of the zero modes, inspired by the 'unpaired electron' idea of Ref.~\cite{ReadRezayi_ZeroModes}.

\subsection{Fixing the Momentum}

\subsubsection{$K_y$-Momentum}

The demonstration of Sec.~\ref{sec:MPS_Torus} can be extended to the spinful case straightforwardly~\cite{OurPaper_HalperinMPS}. The charge sectors $a\in \{0,1/2,1,3/2\}$ are invariant under action of the spin-up and spin-down electronic operators, we can thus define:
\begin{equation} \label{eq:HR_Torus_ConfBlock}
\begin{split}
\Psi_a^{\rm HR} & (w_1, \cdots , w_{N}, w_{[1]}, \cdots , w_{[N]}) = \\ 
& \Tr_a \left[ X \mathcal{V}^\uparrow(w_1) \mathcal{V}^\downarrow(w_{[1]}) \cdots \mathcal{V}^\uparrow(w_{N}) \mathcal{V}^\downarrow(w_{[N]}) \right] \, ,
\end{split}
\end{equation} 
where $P_a$ projects on the states of Eq.~\ref{eq:CFT_Hilbert} with U(1)-charge $q=a$ (mod 2) and the operator $X$ has the form given in Eq.~\ref{eq:X_OperatorsEndTorus}. As in Sec.~\ref{sec:MPS_Torus}, we use the dilatation $\mathcal{V}(w+\alpha L_1) = e^{2i\pi \alpha L_0} \mathcal{V}(w) e^{-2i\pi \alpha L_0}$ and the bosonic commutation relations to derive the effect of $T_1^{\rm cm}$:
\begin{equation}
\begin{split}
\Psi_a^{\rm HR} & \left(w_1 + \frac{L_1}{N_\phi}, \cdots , w_{[N]} + \frac{L_1}{N_\phi}\right) = \\
& \qquad \Tr_a \left[ e^{2i\pi \sqrt{\nu} a_0 } X \mathcal{V}^\uparrow(w_1) \cdots \mathcal{V}^\downarrow(w_{[N]}) \right] \, . 
\end{split}
\end{equation} The operator $e^{2i\pi \sqrt{\nu} a_0 }$ is constant on the charge sector $a$ selected by the projector, which leads to the simple action:
\begin{equation}
	T_1^{\rm cm} \ket{\Psi_a \rangle} = e^{2i\pi \nu a} \ket{\Psi_a \rangle} \, .
\end{equation} 
This proves that specifying the charge sector $a$ corresponds to a many-body quantum number $K_y = 2 a \pi$ in the full Brillouin zone.

\begin{widetext}

\subsubsection{$K_x$-Momentum}

The derivation of a MPS version of Eq.~\ref{eq:HR_Torus_ConfBlock} follows straightforwardly from the study of Sec.~\ref{sec:MPS_Torus}:
\begin{equation} \label{eq:MPS_XAR_TorusHR}
\ket{\Psi_a^{{\rm HR}, \pi} \rangle} = \sum_{ \{(m_j^\uparrow,m_j^\downarrow)\}_{0 \leq j < N_\phi} } \Tr_a \left[X_{\rm AR} \prod_{j=0}^{N_\phi -1} B_{\mathbb{T}}^{(m_j^\uparrow, m_j^\downarrow)} \right] \ket{(m_0^\uparrow,m_0^\downarrow), \cdots (m_{N_\phi-1}^\uparrow,m_{N_\phi-1}^\downarrow) \rangle} \, ,
\end{equation} with
\begin{equation}
B_{\mathbb{T}}^{(m^\uparrow,m^\downarrow)} = \dfrac{1}{\sqrt{m^\uparrow! m^\downarrow!}} U \left( \mathcal{V}_0^\uparrow \right)^{m^\uparrow} \left( \mathcal{V}_0^\downarrow \right)^{m^\downarrow} \, ,
\end{equation} and where $U$ and $X_{\rm AR}$ were defined in Eq.~\ref{eq:Torus_ConformalBlocks} and Eq.~\ref{eq:X_OperatorsEndTorus}. However, it can be seen that this form only produce $K_x = \pi$ eigenvectors. Indeed, consider first the effect of a many-body translation $(T_2^{\rm cm})^2$ on a many-body state:
\begin{equation}
	(T_2^{\rm cm})^2 \ket{\{(m_j^\uparrow,m_j^\downarrow)\}_{0 \leq j < N_\phi} \rangle} = \epsilon  \ket{(m_{N_\phi-2}^\uparrow,m_{N_\phi-2}^\downarrow), (m_{N_\phi-1}^\uparrow,m_{N_\phi-1}^\downarrow), (m_0^\uparrow,m_0^\downarrow), \cdots (m_{N_\phi-3}^\uparrow,m_{N_\phi-3}^\downarrow) \rangle} \, ,
\end{equation} which is inferred from the lowest LL WFs properties (see Eq.~\ref{eq:LLL_Torus_t1}) and where $\epsilon$ is a sign accounting for the reordering of the many-body state. We can use the invariance of the trace under cyclic permutations and the commutation properties of $B_{\mathbb{T}}$ with $X_{\rm AR}$ and $P_a$ to rearrange the MPS matrices in Eq.~\ref{eq:MPS_XAR_TorusHR} in the same order. The commutation of the electronic zero modes cancels out the sign factor $\epsilon$ and $P_a$ is left unchanged by $U^2$, as explained previously. The only non-trivial phase comes from the commutation of $U^2$ with $X_{AR}$ and leads to a factor $(e^{i \pi (N_e-1)/2})^2=-1$. We finally obtain:
\begin{equation}
	\left(T_2^{\rm cm} \right)^2 \ket{\Psi_a^{{\rm HR}, \pi} \rangle}  = e^{i\pi} \ket{\Psi_a^{{\rm HR}, \pi} \rangle} \, .
\end{equation}

To obtain the $K_x=0$ eigenstates, we should note that $X_{\rm AR}$ is not the only way to account for the fermionic anticommutation relations. We could have also used 
\begin{equation}
	X_{\rm F} = e^{ 2 i\pi (N_e-1) G_0^z} \, ,
\end{equation} where $2G_0^z = \sum_{s\neq 0} \psi_{-s}^\dagger \psi_s - \psi_{-s} \psi_s^\dagger$ counts the number of Dirac fermions (recall that the fermionic modes are integer in \textbf{P} and half-integer in \textbf{AP}, see Sec.~\ref{ssec:CFT_HilbertSpace}). $X_{\rm F}$ is the Dirac fermion parity which anticommutes with the electronic operators $\mathcal{V}^\uparrow$ and $\mathcal{V}^\downarrow$. It accomplishes the same purpose as $X_{\rm AR}$ but commutes with $U^2$. The reasoning above applies to the MPS state 
\begin{equation} \label{eq:MPS_XF_TorusHR}
\braket{\{(m_j^\uparrow,m_j^\downarrow)\}_{0 \leq j < N_\phi}}{\Psi_a^{{\rm HR}, 0} \rangle} = \Tr_a \left[X_{\rm F} \prod_{j=0}^{N_\phi -1} B_{\mathbb{T}}^{(m_j^\uparrow, m_j^\downarrow)} \right] \, ,
\end{equation} and shows that it is a $K_x=0$ eigenstate, \textit{i.e.} $(T_2^{\rm cm} )^2 \ket{\Psi_a^{{\rm HR}, 0} \rangle}  = \ket{\Psi_a^{{\rm HR}, 0} \rangle}$.

\end{widetext}

\subsubsection{The Hierarchy Ground States}

The four MPS ansatz built on the HR electronic operators $\{ \Psi_0^{{\rm HR}, 0} , \, \Psi_0^{{\rm HR}, \pi} , \, \Psi_{1/2}^{{\rm HR}, 0} , \, \Psi_{1/2}^{{\rm HR}, \pi} \}$ appear at the position of the zero energy states in ${\rm BZ}_{\rm red}$. They exactly match (up to numerical accuracy) the ED zero energy states for system sizes up to $N_e=10$ particles, which strongly support our derivation. These four WFs were expressed in terms of Weierstrass's elliptic functions in Ref.~\cite{ReadRezayi_ZeroModes}. The latter are essentially determined by the singular part of their behavior near the poles, which are specified by our electronic operators (see Sec.~\ref{sec:HR_Construction}), and by the periodicities which we tuned with the operators $P_a$, $X_{\rm AR}$ and $X_{\rm F}$. It gives a more intuitive way to understand our derivation. The topological sectors are identified by the projectors onto states with an even or odd number of fermions $P_\pm = \frac{1}{2} (X_{\rm AR} \pm X_{\rm F})$~\cite{OurPaper_HalperinMPS}, and the Minimally Entangled States (MES)~\cite{MinimallyEntangledStates} are obtained as linear superpositions of $\Psi_a^{{\rm HR}, 0}$ and $\Psi_a^{{\rm HR}, \pi}$.

The eight zero energy states that we have constructed in the full Brillouin zone (four in the reduced Brillouin zone) are the one which are expected from a level-2 hierarchy state with electronic operator $\mathcal{V}^\uparrow$ and $\mathcal{V}^\downarrow$. The role of zero modes is irrelevant for them, as they can be obtained with the GL representation Eq.~\ref{eq:GuruLudwig_Elec} too, \textit{i.e.} the "small algebra" of Kausch~\cite{Kausch_CuriositiesAtCminu2}. We now proceed to a careful treatment of the zero modes to obtain the two remaining elements of the HR ground state manifold.

\subsection{Ten-Fold Degeneracy and Zero Modes} \label{ssec:TenFold_Degenerate}

\subsubsection{Construction}

We now focus on the $\vec{K} = (0,0)$ momentum where the fifth zero energy state $\widetilde{\Psi}^{\rm HR}$ of the hollow-core Hamiltonian within ${\rm BZ}_{\rm red}$ lies. As shown in the last section, it requires to pick the charge sector $a=0$ in the torus conformal blocks (see Eq.~\ref{eq:Torus_ConformalBlocks}) and to use the operator $X_{\rm F}$ to encode for the fermionic anticommutation relations. This last zero energy state is somehow peculiar as it exhibits some long-range behaviors~\cite{Seidel_ThinTorusForHaldaneRezayi}. Its first quantized form was derived in Ref.~\cite{ReadRezayi_ZeroModes} and can be recast as follows. Let us denote the pairing (or neutral) part of $\ket{\Psi_0^{{\rm HR}, 0}\rangle}$ for a system of $N_e=2N$ particles as $\Delta_{N} ( \{z_i\}_{i=1 \cdots N_e} )$. As in Sec.~\ref{sec:HR_Construction}, we will use bracketed indices $[i]=N+i$ to shorten the notations. We have~\cite{ReadRezayi_ZeroModes}:
\begin{align} \label{eq:Torus_FifthState}
\widetilde{\Psi}^{\rm HR} & \left( \{w_i\}_{i=1 \cdots N_e} \right) = \prod_{i<j} \theta_1 \left(\frac{w_i-w_j}{i L_1} \Big| \tau \right)^2  \\
& \qquad \qquad \sum_{k=1}^N \sum_{\sigma \in \mathfrak{S}_N} \varepsilon(\sigma) \Delta_{N-1} \left( \{w_i\}_{i} \setminus \{ w_{\sigma(1)}, w_{[k]} \}  \right) \, . \notag
\end{align}
The product of Jacobi's theta function $\theta_1$ (see App.~\ref{app:LaughlinWFonTorus}) is nothing but the usual Jastrow factor on the torus~\cite{Fremling_HallViscosityTorus}. The neutral part of Eq.~\ref{eq:Torus_FifthState} can be physically pictured as hosting two unpaired electrons, one spin up at position $w_{\sigma(1)}$ and one spin down at $w_{[k]}$ which obey $\langle \Psi^\sigma \Psi^\rho \rangle = \varepsilon^{\rho \sigma}$ instead of $\langle \Psi^\sigma (z) \Psi^\rho (0)\rangle = \varepsilon^{\rho \sigma}/z^2$ (see Sec.~\ref{ssec:OurCFTDescription}). The sums over $k$ and $\sigma$ antisymmetrize the WF over all possible ways to remove the two electrons at $w_{\sigma(1)}$ and $w_{[k]}$ from the pairing function, which thus only acts on the reduced set $\{w_i\}_{i} \setminus \{ w_{\sigma(1)}, w_{[k]} \}$.

To reproduce Eq.~\ref{eq:Torus_FifthState}, it is crucial to carefully account for the fermionic zero modes~\cite{GurarieFlohrNayak_HaldaneRezayiCFT}. Indeed, they were discarded from the discussion about the mapping of the $c=-2$ symplectic fermion CFT to the unitary $c=1$ Dirac CFT in Ref.~\cite{CapelliTodorov_PairedStates}, which lead to identify the HR theory with that of an Abelian level-2 hierarchy state with eight degenerate ground states on the torus. Such a crude approximation contradicts both analytical results on the exponential number of distinct $2n$ quasiholes states~\cite{ReadRezayi_ZeroModes} and the numerically observed ten-fold GS degeneracy on genus one surfaces~\cite{HermannsRegnaultArdonne_Nonunitaries} (see Sec.~\ref{ssec:Overview_HR}). We now show how we can incorporate the zero modes $\Psi_0^\uparrow$ and $\Psi_0^\downarrow$ in our formalism to exactly reproduce the last GS of the hollow-core model. We note that a non-zero $\Psi_0^\downarrow$ in Eq.~\ref{eq:Full_ElecOperators} implies logarithmic terms in the mode expansion of the fermionic field $\psi$. These corrections are necessary to complete the operator correspondence of the logarithmic $c=-2$ theory~\cite{FlohrLectureLogarithmicCFT} to the $c=1$ theory that we use.

The first consequence of introducing such zero modes is the somehow unusual highest-weight degeneracy in the \textbf{P} sector of the CFT (see Sec.~\ref{ssec:CFT_HilbertSpace}). We have four highest-weight states $\{\ket{\sigma_1}, \ket{\sigma_2}, \ket{\sigma_3}, \ket{\sigma_4} \}$ inherited from the symplectic fermion theory~\cite{Kausch_SymplecticFermions}, which split the CFT Hilbert Space into four blocks. As shown by the chosen computational basis Eq.~\ref{eq:CFT_Hilbert}, the action of the fermionic modes $\Psi_n^\sigma$ with $n \neq 0$ and $\sigma \in \{\uparrow, \downarrow\}$ is block diagonal. We can represent the zero modes and account for their anticommutation relation as:
\begin{equation} \label{eq:ZeroModesJordanForm}
\begin{split}
\Psi_0^\uparrow = 
\kbordermatrix{  & \ket{\sigma_1} & \ket{\sigma_2} & \ket{\sigma_3} & \ket{\sigma_4} \cr
	\ket{\sigma_1}     & 0 & 0 & 0 & 0 \cr
	\ket{\sigma_2}     & \mathds{1} & 0 & 0 & 0 \cr
	\ket{\sigma_3}     & 0 & 0 & 0 & 0 \cr
	\ket{\sigma_4}     & 0 & 0 & \mathds{1} & 0 } 
\, , \\ 
\Psi_0^\downarrow = 
\kbordermatrix{  & \ket{\sigma_1} & \ket{\sigma_2} & \ket{\sigma_3} & \ket{\sigma_4} \cr
	\ket{\sigma_1}     & 0 & 0 & 0 & 0 \cr
	\ket{\sigma_2}     & 0 & 0 & 0 & 0 \cr
	\ket{\sigma_3}     & \mathds{1} & 0 & 0 & 0 \cr
	\ket{\sigma_4}     & 0 & -\mathds{1} & 0 & 0 } 
\, .
\end{split}
\end{equation} We can understand these expressions thanks to the unpaired electron picture. Starting in the sector of the highest weight $\ket{\sigma_1}$, we end up in $\ket{\sigma_4}$ once we have chosen one and exactly one pair of electrons with opposite spins and have left them unpaired since the zero modes act as the identity. All other fermionic modes act identically on the different $\{ \ket{\sigma_i} \}_i$ sectors of the CFT Hilbert space (see Eq.~\ref{eq:CFT_Hilbert}), and all other electrons combine to form the factor $\Delta_{N-1}$ in Eq.~\ref{eq:Torus_FifthState}. 

Introducing the shift operator
\begin{equation}
	P_{14} = \sum_{q, \mu, \eta, \nu} \ket{q,\sigma_1, \mu,\eta,\nu} \bra{q,\sigma_4, \mu,\eta,\nu} \, ,
\end{equation} 
the fifth GS of the hollow-core Hamiltonian in ${\rm BZ}_{\rm red}$ may be written in a MPS form as:
\begin{equation} \label{eq:MPS_Form_FifthGS}
\begin{split}
& \braket{\{(m_j^\uparrow,m_j^\downarrow)\}_{0 \leq j < N_\phi}}{\widetilde{\Psi}^{\rm HR} \rangle} = \\
& \qquad \qquad \qquad \qquad \Tr_0 \left[ P_{14} X_{\rm F} \prod_{j=0}^{N_\phi -1} B_{\mathbb{T}}^{(m_j^\uparrow, m_j^\downarrow)} \right] \, .
\end{split}
\end{equation} We numerically checked that, up to machine precision, the MPS $\ket{\widetilde{\Psi}^{\rm HR} \rangle}$ and $\ket{\Psi_0^{{\rm HR}, 0} \rangle}$ span the whole GS manifold at $\vec{K} = (0,0)$ for up to $N_e=10$ particles, which provide a stringent test of our construction. Because of the shift operator $P_{14}$, the state Eq.~\ref{eq:MPS_Form_FifthGS} should be extracted from a Jordan block of the transfer matrix and it does not appear when diagonalizing the transfer matrix with iterative solvers. This explains the eight degenerate leading eigenvectors of the transfer matrix observed in Sec.~\ref{sec:HR_Cylinder}.

\subsubsection{Characterization}

To summarize the previous construction, we inherit the highest-weight four-fold degeneracy in \textbf{P} from the logarithmic $c=-2$ theory. Only the zero modes connect the different module of the CFT Hilbert space, as described by Eq.~\ref{eq:ZeroModesJordanForm}. Their action simply transcribes in the MPS language the different ways to choose two electrons in an antisymmetric fashion, and resembles long-range matrix product operators. The effect of logarithmic terms in the fermionic mode expansion is manifest in a Jordan block for the largest eigenvalue of the transfer matrix. 

On the torus, the shift operator $P_{14}$ allows to probe the fifth and last representative of the HR GS manifold in ${\rm BZ}_{\rm red}$. These states are not specific to the torus and also appear on zero-genus surfaces, such as the cylinder or the sphere, albeit as quasihole excitations of the densest ground state at magnetic flux $N_\phi = 2N - 2$. To obtain them in finite-size, we simply replace the trace of Eq.\ref{eq:MPS_Form_FifthGS} by left and right MPS boundary states $\ket{0,\sigma_1,0,0,0}$ (or $\ket{1,\sigma_1,0,0,0}$ for its partner in the full Brillouin zone). On the cylinder, we checked that they indeed appear in the zero energy subspace of the hollow-core Hamiltonian at the indicated shift, and are necessary to reproduce all the observed quasihole states. 

\begin{figure}
	\centering
	\includegraphics[width=\columnwidth]{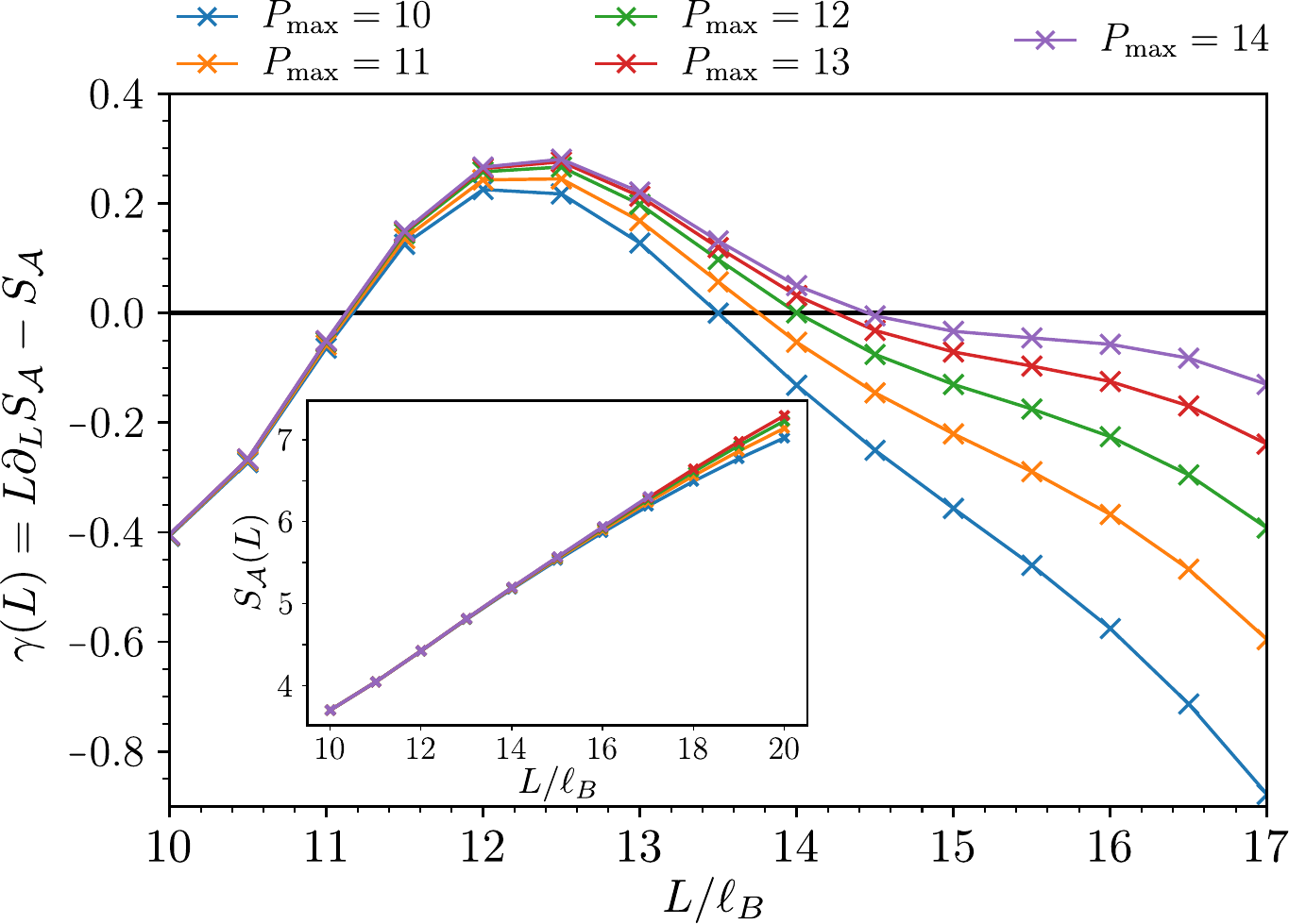}
	\caption{\emph{Numerical extraction of the entanglement entropy $S_{\mathcal{A}}$ for a finite cylinder of length $\sim 120 \ell_b$ cut into two halves for the state $\widetilde{\Psi}^{\rm HR}$ which we could not resolved when diagonalizing the transfer matrix. As in Fig.~\ref{Fig:EntanglementEntropy_iMPS}, the numerical results are plagued with large finite size effects, and the finiteness of the cylinder also introduce another systematic uncertainty. However, the numerically extracted points seem to oscillate around the value $\gamma=0$.}}
	\label{Fig:EntanglementEntropy_FiniteSize}
\end{figure}

This MPS expression allows to characterize the states which we could not resolve when diagonalizing the transfer matrix in Sec.~\ref{sec:HR_Cylinder}. We considered a long but finite cylinder, of length $\sim 120 \ell_B$, and computed the RSEE $S_{\mathcal{A}}^{\rm Fin.}$ of the state $\widetilde{\Psi}^{\rm HR}$ for a cut into two halves. The results are depicted in Fig.~\ref{Fig:EntanglementEntropy_FiniteSize}. We checked that the finite size calculations agree with the iMPS results for the eight GS of studied in Sec.~\ref{sec:HR_Cylinder} (see App.~\ref{app:AdditionalResults}). As in our iMPS calculations, no deviation to the area could be detected over the perimeter range considered. Intriguingly, the first correction to the area law seems to converge toward zero in the thermodynamic limit. Using a different extraction method, we obtain a value slightly away from but still consistent with zero ($\simeq 0.04$, see App.~\ref{app:AdditionalResults}). This behavior is usually encountered in phases with trivial topological order. As a comparison, the same type of calculation for the integer quantum Hall for which the TEE should be strictly equal to zero, typically gives a numerical value of the order of 0.01 for similar perimeter range and  RSEE convergence.

At this stage we do not know how to interpret the apparent lack of constant correction to the area law in the state $\widetilde{\Psi}^{\rm HR}$. Indeed, its usual interpretation in the usual language of quantum dimensions and universal TEE seems moot. The logarithmic nature of the underlying CFT~\cite{Kausch_CuriositiesAtCminu2,Kausch_SymplecticFermions} prevents the identification of a topological charge/sector associated to this state. The potential connection between logarithmic CFTs and topological quantum field theories goes beyond the scope of this paper.

\section{Conclusion}

In this article, we have used the non-unitary CFT description of the Haldane-Rezayi FQH state to derive an exact MPS representation of the latter. A careful treatment of the zero modes enables us to perfectly reproduce the ten ground states of the Haldane-Rezayi phase on the torus obtained with ED, for sufficiently large MPS bond dimensions. There, the non-unitarity of the CFT manifests itself in a Jordan block for the leading eigenvalue of the transfer matrix. Using the MPS techniques, we have shown that the Haldane-Rezayi state has a diverging correlation length in the thermodynamic limit, proving that it does not describe a gapped phase. 

We have also considered the entanglement properties of the Haldane-Rezayi state. For a cylinder with a finite perimeter, we do not see any obvious deviation to the area law. More interestingly, the topological entanglement entropy in all topological sectors seems to only depend on the quantum dimensions of the Abelian excitations. This was already observed for the Gaffnian state, hinting toward a possible generic feature of FQH model states built from non-unitary CFTs. Even more remarkable, the sectors arising from the Jordan block structure do not exhibit any constant corrections, like a topologically trivial state. Future works will try to provide some understanding of this non-unitary state strange behavior.

\section{Acknowledgement} 

We thank A. Gromov and P. Lecheminant for enlightening discussions. We also acknowledge the useful comments and insights of M. Hermanns and E. Ardonne on related topics. VC, BE and NR were supported by the grant ANR TNSTRONG No. ANR-16-CE30-0025 and ANR TopO No. ANR-17-CE30-0013-01.

\newpage

\appendix

\section{SU(2)-Symmetry of the HR State} \label{app:SU2SymmetryHR}

For the sake of simplicity, we will avoid the treatment of all special cases coming from the fermionic zero modes and focus on the \textbf{AP} sector. To show the spin-singlet nature of the HR state, it is sufficient to prove that $S^- \ket{\Psi^{\rm HR}} = S^z \ket{\Psi^{\rm HR}} = 0$. 

Within the bosonized picture, $\psi^\dagger = :e^{i\varphi_s}:$ and $\psi =:e^{-i\varphi_s}:$ with $\varphi_s$ a chiral massless boson with unit compactification radius, it is simple to see why we have $S^z \ket{\Psi^{\rm HR}} = 0$. Indeed, neutrality within the CFT correlator of Eq.~\ref{eq:HR_FirstQuant_AntisymSpin} with respect to the U(1)-charge of $\varphi_s$ ensures that only the spin configurations with equal number of spin up and spin down have non-vanishing MPS coefficients. We will nevertheless exemplify another method to prove this result, which can be generalized to other situations~\cite{OurPaper_HalperinMPS}. We introduce the current
\begin{equation}
	G^z (x) = \frac{1}{2} : \psi^\dagger \psi : (x) = \frac{i}{2} \partial \varphi_s (x) \, .
\end{equation} 
From its OPEs with the electronic operators, we see that its zero mode measures the spin of the electronic operators:
\begin{equation}
(G_0^z \mathcal{V}^\uparrow)(w) = \frac{1}{2} \mathcal{V}^\uparrow(w) \,\, , \,\,\,\, (G_0^z \mathcal{V}^\downarrow)(w) = - \frac{1}{2} \mathcal{V}^\downarrow(w) \, .
\end{equation}
As a consequence, the action of $S^z$ of the HR state of Eq.~\ref{eq:HR_FirstQuant_AntisymSpin} can be described as 
\begin{equation}
S^z \ket{\Psi^{\rm HR}} = \sum_{i=1}^{2N}  \langle \mathcal{O}_{\rm Bkg} \mathcal{W}(w_1) \cdots (G_0^z \mathcal{W})(w_i) \cdots \mathcal{W}(w_{2N}) \rangle \, .
\end{equation}
We evaluate this last formula as follows. Since $G^z$ has conformal dimension one, the correlator $ \langle \mathcal{O}_{\rm Bkg} G^z (x) \prod_i \mathcal{W}(w_i) \rangle $ decays as $1/x^2$ at large distances $|x| \to \infty$. The OPEs with the electronic operators furthermore inform us that:
\begin{widetext} \begin{align} \label{eq:GzCorrelatorFull}
& \left\langle \mathcal{O}_{\rm Bkg} G^z (x) \prod_{i=1}^{2N} \mathcal{W}(w_i) \right\rangle  = \\ 
& \qquad \sum_{i=1}^{2N} \dfrac{1}{x-w_i}  \langle  \mathcal{O}_{\rm Bkg} \mathcal{W}(w_1) \cdots (G_0^z \mathcal{W})(w_i) \cdots \mathcal{W}(w_{2N}) \rangle + \sum_{i=1}^{2N} \dfrac{1}{(x-w_i)^2}  \langle  \mathcal{O}_{\rm Bkg} \mathcal{W}(w_1) \cdots (G_1^z \mathcal{W})(w_i) \cdots \mathcal{W}(w_{2N}) \rangle \, , \notag
\end{align} \end{widetext} with $(G_1^z \psi) (w) = 0$ and $(G_1^z \partial \psi) (w) = -(1/2) \psi$ are the other singular contributions arising in the OPEs. The $1/x$ contribution to Eq.~\ref{eq:GzCorrelatorFull} must be zero because of its long distance behaviour. We conclude that $S^z \ket{\Psi^{\rm HR}} = 0$.

We use a similar argument for the spin lowering operator. We consider the spin-2 field:
\begin{equation}
	G^- (x) = - \frac{1}{2} : \psi \, \partial \psi : = \dfrac{1}{2} : e^{ - 2 i \varphi_s (x) } : \, ,
\end{equation} 
whose OPEs with the electronic operators read:
\begin{subequations} \label{eq:AppSU2_OPEGminus} \begin{align} 
G^- (x) \mathcal{V}^\downarrow (w) & = {\rm reg} \\
G^- (x) \mathcal{V}^\uparrow (w) & = \dfrac{1}{2} \dfrac{\psi (w) \cdot \mathbf{V}^c (w) }{(x-w)^2} + \dfrac{ \mathcal{V}^\downarrow (w) }{x-w} + {\rm reg}
\end{align} \end{subequations} where '${\rm reg}$' denote non-singular terms. The least singular terms of Eq.~\ref{eq:AppSU2_OPEGminus} lead to
\begin{equation}
(G_1^- \mathcal{V}^\uparrow)(w) = \mathcal{V}^\downarrow(w) \,\, , \,\,\,\, (G_1^- \mathcal{V}^\downarrow)(w) = 0  \, ,
\end{equation} and allow to map the action of $S^-$ in the CFT as:
\begin{equation} \label{eq:AppSU2_Sminus}
S^- \ket{\Psi^{\rm HR}} = \sum_{i=1}^{2N}  \langle \mathcal{O}_{\rm Bkg} \mathcal{W}(w_1) \cdots (G_1^- \mathcal{W})(w_i) \cdots \mathcal{W}(w_{2N}) \rangle \, .
\end{equation} Although $G^- (x)$ is not a usual current, it has conformal dimension two. As a consequence, the correlator $ \langle \mathcal{O}_{\rm Bkg} G^- (x) \prod_i \mathcal{W}(w_i) \rangle $ decays as $1/x^4$ at large distances. Its $1/x^3$ contribution which exactly matches Eq.~\ref{eq:AppSU2_Sminus} must be zero, which shows that $S^- \ket{\Psi^{\rm HR}} = 0$.

\section{Holomorphic Structure of the Lowest LL on Torus} \label{app:HolomorphicLLL}

We derived the form of the LLL one-body WFs $\phi (z) = e^{- x^2 / (2 \ell_B^2)} f(w)$ in Eq.~\ref{eq:LLL_Torus_GeomFactors}, where $f$ is a holomorphic function. In this appendix, we look more closely at the properties of these holomorphic sections on the torus pierced by $N_\phi$ flux quanta.

We first show that the LLL has dimension $N_\phi$. Consider the auxiliary function $g = (\log \circ f)' = f'/f$. It has a simple pole for each zero of $f$, each with residue equal to one. The contour integral of $g$ around the torus is thus equal to the number of zero of $f$ in the torus' principal region. Thanks to the boundary conditions satisfied by $g$:
\begin{equation}
g(w+iL_1) = g(w) \, , \, \, g(w+i e^{i\theta} L_2) = g(w) - 2 \pi \frac{N_\phi}{L_1} \, ,
\end{equation} we can compute the contour integral directly and get that the number of zeros of $f$ is $N_\phi$. Once the zeros are set to given positions $\{ w_j \, | j = 1 , \cdots , N_\phi \}$, the function $f$ is almost completely specified:
\begin{equation}
f(w) = e^{k w} \prod_{j=1}^{N_\phi} \theta_1 \left(\frac{w-w_j}{iL_1} \Big| \tau \right) \, ,
\end{equation} 
with the following constraints deriving from the TBC:
\begin{equation} \label{eq:App_ConditionSolutionLLL}
e^{i k L_1} = e^{\frac{2 \pi}{L_1} \sum_j w_j + i k L_1 \tau} = (-1)^{N_\phi} \, .
\end{equation} 
The Riemann-Roch theorem states that there are $N_\phi$ linearly independent solutions to these conditions, which form a basis of the LLL. In the main text, we give a LLL basis made of $t_1$ eigenvectors. It is obtained by placing the zeros equally space on a vertical line. More precisely for $k_y = 0, \cdots , N_\phi-1$ we choose 
\begin{equation}
	w_j (k_y) = w_0 (k_y) + i j L_1/N_\phi \, , \quad j = 1, \cdots , N_\phi -1  \, ,
\end{equation} with
\begin{equation}
	\dfrac{N_\phi}{i L_1} w_0 (k_y) = \dfrac{1}{2} + N_\phi \tau \left( \dfrac{1}{2} - \dfrac{k_y}{N_\phi} \right) \, .
\end{equation} One can check that this choice is consistent with the TBC and satisfy Eq.~\ref{eq:App_ConditionSolutionLLL} with $kL_1 = - \pi N_\phi + 2 \pi k_y$. The function
\begin{equation*}
f_{k_y}(w) = e^{(2 \pi k_y - \pi N_\phi) w /L_1} \prod_{j=1}^{N_\phi} \theta_1 \left(\frac{w-w_j(k_y)}{iL_1} \Big| \tau \right) 
\end{equation*} thus satisfy the TBC and possesses the same zeros as the function of Eq.~\ref{eq:LLL_Torus_t1} (see App.~\ref{app:Elliptic_Function}). They can be identified up to an irrelevant constant factor:
\begin{equation}
f_{k_y}(w) = \dfrac{1}{\sqrt{L_1 \ell_B \sqrt{\pi}}} \vartheta \begin{bmatrix} k_y/N_\phi \\	0 \end{bmatrix} \left( \frac{N_\phi}{i L_1} w \Big| N_\phi \tau \right) \, .
\end{equation}

\begin{widetext}

\section{The Conformal Blocks Satisfy the TBC}\label{app:TBCforConfBlocks}

We now prove that the choice Eq.~\ref{eq:Torus_ConformalBlocks} indeed leads to the correct quasi-periodic conditions on the torus. We start with $\Psi_a (w_1 + i L_2 e^{i \theta}, w_2 \cdots , w_{N_e})$. Using the fermionic anticommutation relations, we bring $\mathcal{V}(w_1 + i L_2 e^{i\theta})$ to the rightmost part of the trace. Using the invariance of the trace under cyclic permutations and the fact that topological sectors are stable by action of the electronic operator, we get:
\begin{equation}
\Psi_a (w_1 + i L_2 e^{i \theta}, w_2 \cdots , w_{N_e}) = (-1)^{N_e-1} \Tr_a \left[\mathcal{V}(w_1 + i L_2 e^{i\theta})X_{\rm AR} X_{\rm P} X_{\rm Bkg} \mathcal{V}(w_2) \cdots \mathcal{V}(w_{N_e}) \right] \, . 
\end{equation} Since $e^{\alpha \sqrt{\nu} a_0} \mathcal{V}(w) = e^\alpha \mathcal{V}(w) e^{\alpha \sqrt{\nu} a_0}$, the sign factor cancels out when $\mathcal{V}(w_1 + i L_2 e^{i \theta})$ commutes with $X_{\rm AR}$. Dilatations on the plane are generated by $L_0$, and the commutation with $X_{\rm P}$ can be inferred from $q^{L_0} \mathcal{V}(w) q^{-L_0} = \mathcal{V}(w + i L_2 e^{i\theta}) $ with $q=\exp(2i\pi\tau)$. We already treated the case of the background operator $X_{\rm Bkg}$ in Sec.~\ref{sec:HR_Cylinder}, thanks to $e^{-i\sqrt{\nu} \varphi_0} \mathcal{V}(w) = z \mathcal{V}(w) e^{-i\sqrt{\nu} \varphi_0}$. Combining the different pieces, we end up with:
\begin{equation}
\Psi_a (w_1 + i L_2 e^{i \theta}, w_2 \cdots , w_{N_e}) = \exp\left[ -2i\pi N_\phi \left( \frac{w_1}{i L_1} + \frac{\tau}{2} \right) \right] \Psi_a (w_1 , \cdots , w_{N_e}) \, ,
\end{equation} which is the result expected from the TBC (see Eq.~\ref{eq:MagneticTranslation_BC} and Eq.~\ref{eq:LLL_Torus_GeomFactors}).

To prepare for the derivation of the MPS representation, we note that a similar derivation can be used to get the following identity:
\begin{equation}
\Tr_a \left[X_{\rm AR} X_{\rm P} X_{\rm Bkg} \mathcal{V}_{-\lambda_1} \cdots \mathcal{V}_{-\lambda_i -n N_\phi} \cdots \mathcal{V}_{-\lambda_{N_e}} \right] = q^{n \lambda_1 + n^2 N_\phi/2} \Tr_a \left[X_{\rm AR} X_{\rm P} X_{\rm Bkg} \mathcal{V}_{-\lambda_1} \cdots \mathcal{V}_{-\lambda_i} \cdots \mathcal{V}_{-\lambda_{N_e}} \right] \, .
\end{equation} We have used the mode expansion $\mathcal{V} (w) = \sum_\lambda z^\lambda \mathcal{V}_{-\lambda}$ where we recall the mapping $z = \exp (\frac{2 \pi}{L_1} w )$. Summing the last equation over $n \in \mathbb{Z}$ brings out the torus lowest LL WF of Eq.~\ref{eq:LLL_Torus_t1}:
\begin{equation} \label{eq:MassageModeExpansionForMPS}
\sum_{\lambda_i \in s+N_\phi \mathbb{Z}} \Tr_a \left[X_{\rm AR} X_{\rm P} X_{\rm Bkg} \mathcal{V}_{-\lambda_1} \cdots \mathcal{V}_{-\lambda_{N_e}} \right] w_i^{\lambda_i} = \Tr_a \left[X_{\rm AR} X_{\rm P} X_{\rm Bkg} \mathcal{V}_{-\lambda_1} \cdots \mathcal{V}_{-s} \cdots \mathcal{V}_{-\lambda_{N_e}} \right] e^{-i\pi \tau s^2/N_\phi} f_s(w_i) \, .
\end{equation}

\end{widetext}

\section{Laughlin WFs on the Torus} \label{app:LaughlinWFonTorus}

In this appendix, we show that our construction of Eq.~\ref{eq:Torus_ConformalBlocks} can exactly reproduce the $m$ degenerate GS of the Laughlin phase at filling factor $\nu=1/m$, $m\in \mathbb{N}^*$. Their explicit real-space expression was derived by Haldane and Rezayi in Ref.~\cite{HaldaneRezayi_TorusGeom}:
\begin{equation} \label{eq:LaughlinOnTorus_Litterature}
\Psi_a^{\rm Lgh} = \vartheta \begin{bmatrix} a/m + t \\ - m t \end{bmatrix} \left( \frac{m W}{i L_1} \Big| m \tau \right) \prod_{i<j} \theta_1 \left( \frac{w_i - w_j}{iL_1} \Big| \tau \right)^m \, .
\end{equation} The first $\vartheta$ function only depends on the center-of-mass coordinate $W= \sum_i w_i$, and distinguishes the different GS by their momentum quantum number $K_y = 2a\pi/m $ through the parameter $a = 0, \cdots , m-1 $. We have also introduced $t = (N_e-1)/2$. The product of $\theta_1$ is the usual Jastrow factor which provides the correct vanishing properties to $\Psi_a^{\rm Lgh}$ when two electrons get close to one another.

The underlying CFT for the Laughlin phase does not have any neutral component, and the electronic operator reads
\begin{equation}
\mathcal{V} (w) = :e^{i \sqrt{m} \varphi (w)} :
\end{equation} in which the free chiral boson $\varphi$ is compactified on a circle of radius $R=\sqrt{m}$. Its two-point correlation function is the Green's function of the Laplacian: $\langle \varphi(w) \varphi(0) \rangle = - \log z$. The different topological sector for the Laughlin state are simply charge sectors $a$ gathering all states with U(1)-charge $q = a$ (mod $m$), see Sec.~\ref{ssec:CFT_HilbertSpace}.

We want to show that the conformal block of Eq.~\ref{eq:Torus_ConformalBlocks}, that we denote as $\Psi_a^{\rm CB}$, reproduces Eq.~\ref{eq:LaughlinOnTorus_Litterature}. Using the identity~\cite{YellowBook}
\begin{equation}
\prod_{i=1}^{N_e} \mathcal{V} (w_i) = \prod_{i<j} (z_i-z_j)^m  : e^{i \sqrt{m} \sum_i \varphi(w_i) } : \, , 
\end{equation} we can focus on the trace of the normal ordered operator 
\begin{equation}
\Tr_a [X : \exp(i\sqrt{m}\sum_i \varphi(w_i)) : ] = Q_a \prod_{n \in \mathbb{N}^*} P_n \, ,
\end{equation} 
which naturally decouples the contribution of the different bosonic modes as
\begin{subequations} \begin{align} \label{eq:AppLaughlin_Qa}
	& Q_a = \Tr_a^Q \left[ X_{\rm AR} q^{\frac{a_0^2}{2} + \frac{N_\phi}{2\sqrt{m}} a_0} e^{\frac{2\pi}{L_1} \sqrt{m} W a_0} \right] \, , \\
	& P_n = \Tr^n \left[ q^{a_{-n}a_n} e^{\frac{\sqrt{m}}{n} a_{-n} \sum_i z_i^n} e^{- \frac{\sqrt{m}}{n} a_{n} \sum_i z_i^{-n}} \right] \, .
\end{align} \end{subequations} 
Here the notations $\Tr_a^Q$ and $\Tr^n$ respectively mean a trace over the possible U(1)-charges in topological sector $a$ and the degrees of freedom associated with the $n$-th creation $a_{-n}$ and annihilation $a_{n}$ bosonic modes.

For all $n \in \mathbb{N}^*$, the operators $b^\dagger = a_{-n} / \sqrt{n}$ and $b = a_n / \sqrt{n}$ are the creation and annihilation operators of a harmonic oscillator. Using a coherent state basis, we can derive
\begin{equation}
\Tr \left[ q^{n b^\dagger b} e^{\gamma b^\dagger} e^{\delta b} \right] = \dfrac{1}{1-q^n} \exp\left( \dfrac{\gamma \delta q^n}{1-q^n} \right) \, ,
\end{equation} which allows us to evaluate all the $P_n$'s. This leads us to:
\begin{equation}\label{eq:AppLaugh_Proof1}
\Psi_a^{\rm CB} = \dfrac{Q_a}{\eta (\tau)} e^{\frac{2\pi}{L_1} m W t} \prod_{i<j} {\rm Aux} \left(\dfrac{w_i-w_j}{iL_1}\right)^m \, ,
\end{equation} 
where we have introduced the Dedekind function $\eta(q)$. Up to an inconsequential multiplicative prefactor, the auxiliary function reads:
\begin{align}
& {\rm Aux} \left(\dfrac{w}{iL_1}\right) \\
& = \dfrac{z^{1/2} - z^{-1/2}}{2i\pi} \prod_{p \in \mathbb{N}^*} \exp\left( - \dfrac{q^p (z^{p} - z^{-p} -2)}{p (1-q^p)}\right) \\
& = \alpha \theta_1  \left(\dfrac{w}{iL_1} \Big| \tau \right) \, , \label{eq:AppLaugh_Proof2}
\end{align} with $z = \exp(2\pi w/L_1)$ and $\alpha = 1/\partial_w \theta_1  ( 0 | \tau )$ (see App.~\ref{app:Elliptic_Function}). Hence, the Jastrow part of the Laughlin state Eq.~\ref{eq:LaughlinOnTorus_Litterature} is reproduced by the product of $P_n$. What remains to be computed in the model WF $\Psi_a^{\rm CB}$ is the zero mode contribution $Q_a$ which only depends on the center of mass position.

For the fermionic Laughlin states that we consider, we have $m$ odd. In that case, $X_{\rm AR}$ acts as a real phase factor on the charge basis states Eq.~\ref{eq:CFT_Hilbert} and it can be replaced by $(X_{\rm AR})^m$ in $Q_a$ without changing the state (see Eq.~\ref{eq:AppLaughlin_Qa}). Summing over the allowed charges in topological sector $a$ gives, up to a global phase factor:
\begin{align}
Q_a e^{\frac{2 \pi}{L_1} m W t} & = \!\! \sum_{k \in a/m+\mathbb{Z}} \!\! q^{\frac{m}{2}(k^2 + N_e k)} e^{2i\pi (k + t) (\frac{mW}{iL_1} - m t)} \notag \\
& = \vartheta \begin{bmatrix} a/m + t \\ - m t \end{bmatrix} \left( \frac{m W}{i L_1} + \dfrac{m\tau}{2} \Big| m \tau \right) \, . \label{eq:AppLaugh_Proof3}
\end{align} 
Equations~\ref{eq:AppLaugh_Proof1}~-~\ref{eq:AppLaugh_Proof2}~and~\ref{eq:AppLaugh_Proof3} proves that our approach indeed reproduce the Laughlin states of Eq.~\ref{eq:LaughlinOnTorus_Litterature} on the torus, with a slight difference in the choice of the origin.

\section{Elliptic Functions} \label{app:Elliptic_Function}

The generalized theta function, specified by two real parameters $a$ and $b$, depends on two complex variables $z$ and $\tau$ as:
\begin{equation}
\vartheta \begin{bmatrix} a \\ b \end{bmatrix} \left( w | \tau \right) = \sum_{k \in a + \mathbb{Z}} e^{i \pi \tau k^2} e^{2i\pi k (w+b)} \, .
\end{equation}

Using the Jacobi's triple product identity~\cite[Chap.~10]{YellowBook}, we can see that the zeros are located at 
\begin{equation}
w = \left( \frac{1}{2} - b + m_1 \right) + \tau \left( \frac{1}{2} - a + m_2 \right)
\end{equation} with $m_1, m_2 \in \mathbb{Z}$. This was important when deriving the explicit form of the LLL basis in App.~\ref{app:HolomorphicLLL}.

Other useful formulas when considering the TBC are:
\begin{subequations}
\begin{align}
\vartheta \begin{bmatrix} a +1 \\ b \end{bmatrix} \left( w | \tau \right) & = \vartheta \begin{bmatrix} a \\ b \end{bmatrix} \left( w | \tau \right) \, ,
\\
\vartheta \begin{bmatrix} a \\ b \end{bmatrix} \left( w + 1 | \tau \right) & = e^{2 i \pi a} \vartheta \begin{bmatrix} a \\ b \end{bmatrix} \left( w | \tau \right) \, ,
\\
\vartheta \begin{bmatrix} a \\ b \end{bmatrix} \left( w + \tau | \tau \right) & = e^{-2i\pi (w+b)-i \pi \tau} \vartheta \begin{bmatrix} a + 1 \\ b \end{bmatrix} \left( w | \tau \right) \, .
\end{align}
\end{subequations} They allow to check that the LLL basis Eq.~\ref{eq:LLL_Torus_t1} satisfy the TBC and to compute the effect of $t_1$ and $t_2$ on the latter.

Finally, the last function used in the article is the $\theta_1$ function (see Eq.~\ref{eq:Torus_FifthState}), which is conveniently expressed as: 
\begin{equation}
\theta_1 ( w | \tau ) = - \vartheta \begin{bmatrix} 1/2 \\ 1/2 \end{bmatrix} \left( w | \tau \right) \, .
\end{equation} This function is necessary to describe the Jastrow factors on the torus. However, it requires some work to recast it in the form encountered in App.~\ref{app:LaughlinWFonTorus}. Using Jacobi's triple product identity, we first have~\cite{YellowBook}:
\begin{align}
\theta_1 & ( w | \tau ) = \\ 
& - i y^{1/2} q^{1/8} \prod_{n \in \mathbb{N}^*} (1-q^n) \prod_{n \in \mathbb{N}} (1-y q^{n+1}) (1-y^{-1} q^{n}) \, , \notag
\end{align} with $y=e^{2i\pi w}$ and $q=e^{2i\pi \tau}$. We can use this expression and the serie's expansion $\log (1-z) = -\sum_{p \geq 1} z^p/p$ to get:
\begin{align}
&\frac{\theta_1  ( w | \tau )}{\partial_w \theta_1  ( 0 | \tau )} \\
& = \dfrac{y^{1/2}-y^{-1/2}}{2i\pi} \prod_{n \in \mathbb{N}^*} \dfrac{(1-yq^n)(1-y^{-1}q^n)}{(1-q^n)(1-q^n)} \\
& = \dfrac{y^{1/2}-y^{-1/2}}{2i\pi} \exp \left( \! - \!\! \sum_{p \in \mathbb{N}^*} \dfrac{q^p (y^{p/2}-y^{-p/2})^2}{p (1-q^p)}  \right) \\
& = {\rm Aux} (w) \, .
\end{align}

\section{Additional Numerical Results} \label{app:AdditionalResults}

In this appendix, we provide additional numerical evidence about the anomalous topological entanglement entropy values for the Haldane-Rezayi state.

\subsection{Orbital Entanglement Entropy}

\begin{figure}
	\centering
	\includegraphics[width=\columnwidth]{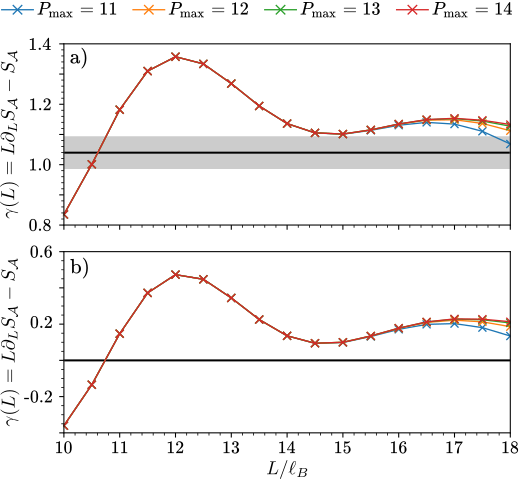}
	\caption{\emph{ Numerical extraction of the OEE for a) a half infinite cylinder in the topological sector $a=1$ with even fermionic parity and b) a finite cylinder of length $\sim 120 \ell_b$ cut into two halves for the state $\widetilde{\Psi}^{\rm HR}$. They should respectively be compared to the RSEE results of Fig.~\ref{Fig:EntanglementEntropy_iMPS} and Fig.~\ref{Fig:EntanglementEntropy_FiniteSize}. The OEE and RSEE results agree, up to few percents discrepancies.}}
	\label{Fig:OES_iMPS}
\end{figure}

The topological entanglement entropy for the Gaffnian state was extracted in Ref.~\cite{RegnaultEstienne_CorrelationLength} with an orbital cut. Rigorously, theoretical results on the area law and its first universal correction only hold true for a real-space cut. Indeed, it is not clear whether other corrections appear for orbital cuts, even though it is believed that both cuts should lead to the same topological entanglement entropy in the thermodynamic limit. To compare both approaches, we have computed the Orbital Entanglement Entropy (OEE) of the states investigated in the article, namely the eight GS accessible in iMPS calculations (see Fig.~\ref{Fig:OES_iMPS}a) and the other two described in Sec.~\ref{ssec:TenFold_Degenerate} (see Fig.~\ref{Fig:OES_iMPS}b). For this latter, we have considered a long, but finite, cylinder. As in Ref.~\cite{OurPaper_HalperinMPS}, we observe that the OEE has a similar behavior as the RSEE although the extracted constant corrections are slightly off by a few percent.

\subsection{Finite Size RSEE}

\begin{figure}
	\centering
	\includegraphics[width=\columnwidth]{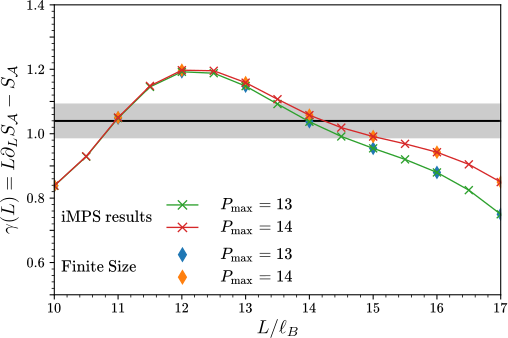}
	\caption{ \emph{ Numerical extraction of the topological entanglement entropy for a finite cylinder of length $\sim 120 \ell_B$ (diamond symbols) cut into two halves in the topological sector $a=0$ with even fermionic parity. The results perfectly agree with the iMPS calculations of Fig.~\ref{Fig:EntanglementEntropy_iMPS} (performed on a finer grid), which we are also shown here ($\times$ symbols). This supports our finite size calculations of Sec.~\ref{ssec:TenFold_Degenerate}.} }
	\label{Fig:CompareFiniteVsIMPS}
\end{figure}

We tested the finite size RSEE calculations of Sec.~\ref{ssec:TenFold_Degenerate} with the hierarchy GS, for which we can assess quantitatively the cylinder finite size effects thanks to the iMPS results of Sec.~\ref{ssec:iMPS_EE}. The two methods, compared in Fig.~\ref{Fig:CompareFiniteVsIMPS}, agree to less than a percent for the subleading correction $\gamma(L)$. This consistency check validates our finite size calculations of Sec.~\ref{ssec:TenFold_Degenerate} for the non-Abelian states.

\subsection{Another Extraction of the TEE}

\begin{figure}
	\centering
	\includegraphics[width=\columnwidth]{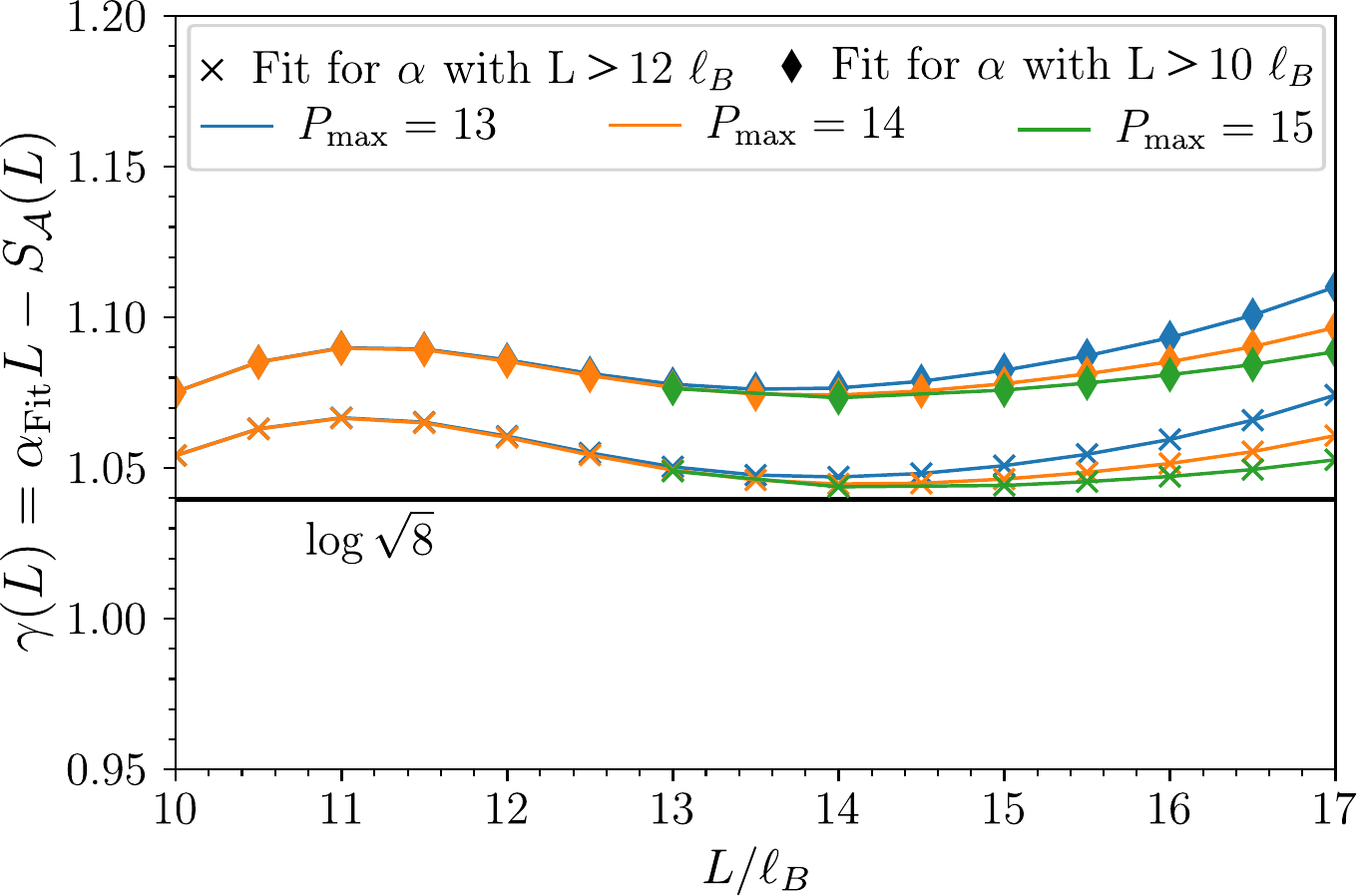}
	\caption{ \emph{Topological Entanglement Entropy $\gamma(L)= \alpha_{\rm Fit} L - S_\mathcal{A}(L)$ with $\alpha_{\rm Fit}$ extracted from a fit on $S_\mathcal{A}(L)$ (obtained in topological sector $a=1$ with even fermionic parity on an infinite cylinder, as in Fig.~\ref{Fig:EntanglementEntropy_iMPS}). The fit is performed over all the points converged with respect to the truncation parameter $P_{\rm max}$ but discards some points at small $L$ to avoid finite size effects. This introduce a selection bias which we exemplify by showing the results for a fit with (diamonds) and without ($\times$ symbols) the points at $L=11 \ell_B$. All results nevertheless agree with the value of $\log\sqrt{8}$ given in the main text. }}
	\label{Fig:OtherFitting}
\end{figure}

Using finite differences on the RSEE data is not the only way to extract the TEE. We also performed linear fits on the RSEE to determine the linear coefficient $\alpha$ to the area law Eq.~\ref{eq:AreaLaw} and subtracted it subsequently (as was used for the Gaffnian state in Ref.~\cite{RegnaultEstienne_CorrelationLength}). Fig.~\ref{Fig:OtherFitting} displays the results of such a procedure for the topological sector $a=1$ with even fermionic parity. We find this approach to average the errors on the extensive part of the RSEE, and thus to give more precise results for the TEE (all equal to $\log\sqrt{8}$ within a few percents). This method is less sensitive to truncation effects but at the same time introduces a selection bias in the points chosen to perform the fit. For instance, we show in Fig.~\ref{Fig:OtherFitting} how the extracted TEE changes when the point at $L=11\ell_B$ is either in or out the selected points for the fit. We have decided to only display the finite difference results in the main text, which seem less precise but already show the correct convergence behaviors.

\begin{figure}
	\centering
	\includegraphics[width=\columnwidth]{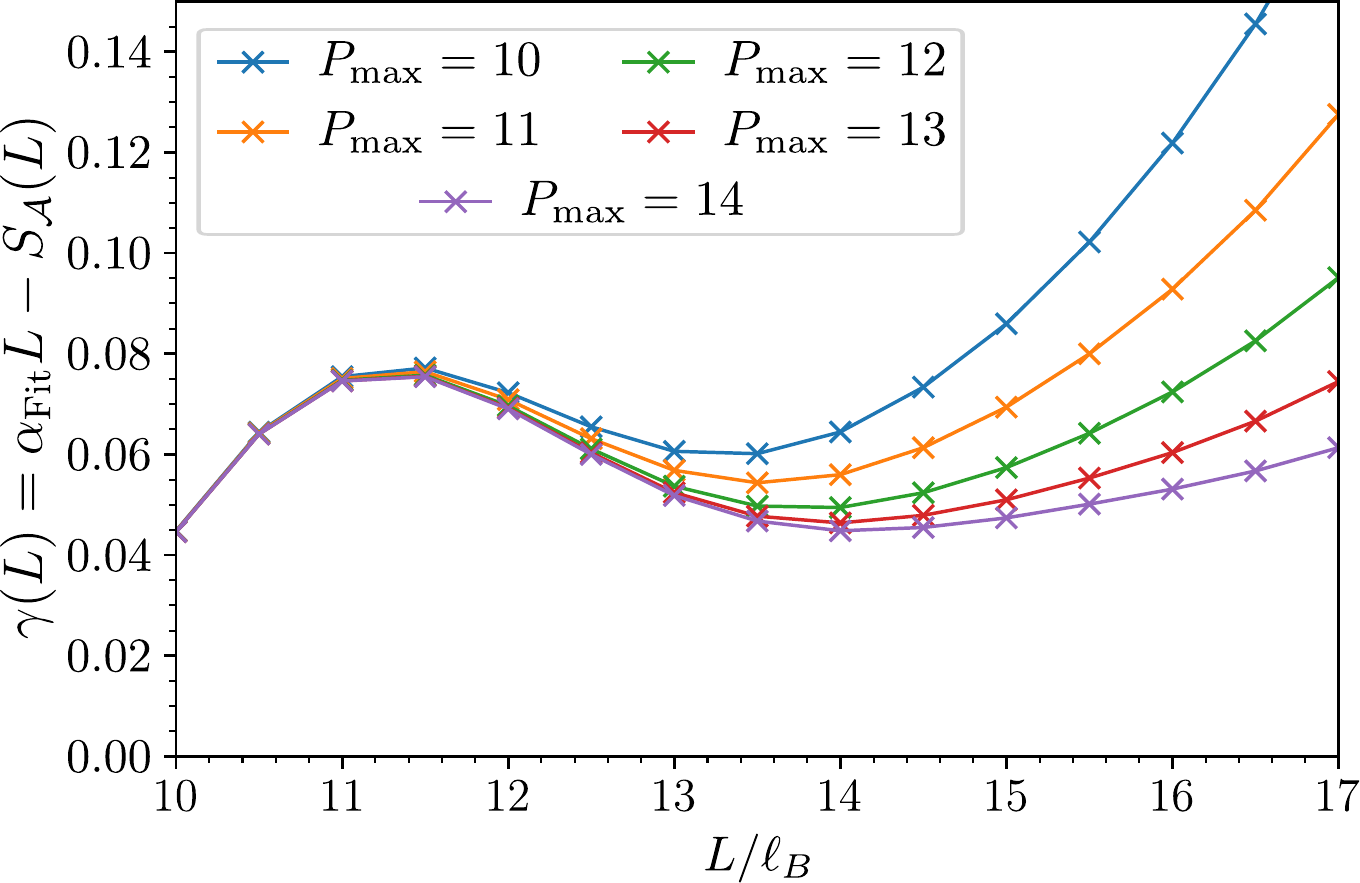}
	\caption{ \emph{Topological Entanglement Entropy $\gamma(L)= \alpha_{\rm Fit} L - S_\mathcal{A}(L)$ with $\alpha_{\rm Fit}$ extracted from a fit on $S_\mathcal{A}(L)$ (obtained for the state $\widetilde{\Psi}^{\rm HR}$ on an finite cylinder, as in Fig.~\ref{Fig:EntanglementEntropy_iMPS}). The fit is performed over all the points converged with respect to the truncation parameter $P_{\rm max}$ but discards the points at $L<12 \ell_B$ to avoid finite size effects. }  }
	\label{Fig:OtherFitting_Weird0}
\end{figure}

We performed a similar analysis for the GS arising from the Jordan block of the transfer matrix, described in Sec.~\ref{ssec:TenFold_Degenerate}. The results are displayed in Fig.~\ref{Fig:OtherFitting_Weird0}. They are slightly away from but still consistent with zero, the value extracted from Fig.~\ref{Fig:EntanglementEntropy_FiniteSize} in the main text.

\newpage


\bibliography{Biblio_HR}

\end{document}